\documentclass{aastex62}
\usepackage{amsmath}
\newcommand\dd{{\mathrm{d}}}

\received{}
\revised{}
\accepted{}

\begin{document}

\title{Active-sterile Neutrino Oscillations in Neutrino-driven Winds:
Implications for Nucleosynthesis}

\author{Zewei Xiong}
\affil{School of Physics and Astronomy, University of Minnesota, Minneapolis, Minnesota 55455, USA}

\author{Meng-Ru Wu}
\affil{Institute of Physics, Academia Sinica, Taipei 11529, Taiwan}
\affil{Institute of Astronomy and Astrophysics, Academia Sinica, Taipei 10617, Taiwan}

\author{Yong-Zhong Qian}
\affil{School of Physics and Astronomy, University of Minnesota, Minneapolis, Minnesota 55455, USA}
\affil{Tsung-Dao Lee Institute, Shanghai 200240, People's Republic of China}

\begin{abstract}
A protoneutron star produced in a core-collapse supernova (CCSN) drives a wind by
its intense neutrino emission. We implement active-sterile neutrino oscillations 
in a steady-state model of this neutrino-driven wind to study their effects on
the dynamics and nucleosynthesis of the wind in a self-consistent manner.
Using vacuum mixing parameters indicated by some experiments for a sterile $\nu_s$ 
of $\sim 1$~eV in mass, we observe interesting features of oscillations due to 
various feedback. For the higher $\nu_s$ mass values, we find that oscillations can 
reduce the mass loss rate and the wind velocity by a factor of $\sim 1.6$--2.7 and 
change the electron fraction critical to nucleosynthesis by a significant to large amount. 
In the most dramatic cases, oscillations shifts nucleosynthesis from dominant production of
$^{45}$Sc to that of $^{86}$Kr and $^{90}$Zr during the early epochs of the CCSN evolution.
\end{abstract}


\section{Introduction}
In this paper we study the effects of active-sterile neutrino oscillations 
on the dynamics and nucleosynthesis of neutrino-driven winds. The proto-neutron 
star (PNS) formed in a core-collapse supernova (CCSN) is a profuse source of 
$\nu_e$, $\bar\nu_e$, $\nu_\mu$, $\bar\nu_\mu$, $\nu_\tau$, and $\bar\nu_\tau$. 
These neutrinos interact with the material in the vicinity of the PNS
mainly through
\begin{subequations}
\begin{eqnarray}
\nu_e+n&\to& p+e^-, \label{eq:nue_neutron_scattering}\\
\bar\nu_e+p&\to& n+e^+. \label{eq:nuebar_proton_scattering}
\end{eqnarray}
\end{subequations}
This heating drives a wind \citep{duncan1986neutrino},
which eventually becomes part of the ejecta from the CCSN. This 
neutrino-driven wind has been studied extensively as a site for production of 
elements heavier than the Fe group \citep{woosley1992collapse,woosley1992alpha,%
meyer1992r,woosley1994r,witti1994nucleosynthesis,takahashi1994nucleosynthesis,%
qian1996nucleosynthesis,hoffman1997nucleosynthesis,wanajo2001r,thompson2001physics,%
roberts2010integrated,wanajo2013r}. The neutrino reactions in 
Eqs.~(\ref{eq:nue_neutron_scattering}) and (\ref{eq:nuebar_proton_scattering}) 
not only provide the heating, but also determine the neutron-to-proton ratio
of the wind \citep{qian1993connection,qian1996nucleosynthesis}. This ratio is
equivalent to the net number of electrons per baryon, i.e., the electron fraction 
$Y_e$, and is a critical parameter for nucleosynthesis. Because neutrinos of 
different flavors interact differently with matter, neutrino flavor oscillations
would potentially impact the dynamics and 
nucleosynthesis of the wind (e.g., \citealt{qian1993connection}).
Here we focus on the oscillations between $\nu_e$ ($\bar\nu_e$) 
and $\nu_s$ ($\bar\nu_s$), where $\nu_s$ ($\bar\nu_s$) is a sterile species that 
does not have normal weak interaction like an active one.

The existence of sterile neutrinos have been discussed based on both 
theoretical and experimental considerations (e.g., \citealt{abazajian2012light}). 
Potential effects of active-sterile
neutrino oscillations on the dynamics and nucleosynthesis of CCSNe
have also been studied (e.g., \citealt{nunokawa1997supernova,mclaughlin1999active,%
tamborra2012impact,wu2014impact,pllumbi2015impact}).
These previous studies, however, are not fully self-consistent in that
they did not take into account the feedback of active-sterile neutrino 
oscillations on the velocity ($v$), density ($\rho$), and temperature ($T$) 
profiles of the CCSN. In this paper we make a significant step towards 
treating such feedback by coupling neutrino flavor evolution with the evolution 
of $v$, $\rho$, $T$, and $Y_e$ in the neutrino-driven wind.

This paper is organized as follows. In \S\ref{sec:setup}, we describe our
steady-state model of the neutrino-driven wind including active-sterile 
neutrino oscillations. In \S\ref{sec:results}, we present the results for
nine cases that correspond to winds ejected at three representative
times of CCSN evolution with three different sets of neutrino oscillation 
parameters. For each case, we show the profiles of $v$, $\rho$, $T$, and $Y_e$ 
in the wind and compare them with those in the absence of oscillations. 
In \S\ref{sec:nucleosynthesis}, we show the effects of active-sterile neutrino 
oscillations on nucleosynthesis in the wind. We summarize our results and give 
conclusions in \S\ref{sec:conclusions}.

\section{Model of the neutrino-driven wind} \label{sec:setup}
A massive star arrives at the end of its life when nuclear fuel is exhausted
in its core. The core undergoes gravitational collapse and bounces when
supranuclear density is reached at the center and a PNS
is born. The bounce launches a shock, which is stalled on the way out of the 
core due to energy loss from dissociating nuclei into mostly free nucleons. 
How this shock is revived to make an explosion is the crux of the CCSN 
mechanism (e.g., \citealt{janka2012explosion}). Neutrinos 
emitted by the PNS are thought to play a critical role by heating the material 
behind the stalled shock mainly through the reactions
in Eqs.~(\ref{eq:nue_neutron_scattering}) and 
(\ref{eq:nuebar_proton_scattering}) \citep{bethe1985revival}.
Because the neutrino emission lasts for $\sim 10$~s, the same heating 
continues to drive a mass outflow, the so-called neutrino-driven wind, 
after the shock is revived.

While the wind can be studied within a CCSN simulation,
its physical conditions can be understood quite well by treating it
as a stationary mass outflow between the PNS and the shock 
\citep{qian1996nucleosynthesis}. In the latter approach, the wind is
the solution to an eigenvalue problem with the inner and outer
boundaries specified by the PNS and shock, respectively. The eigenvalue
to be determined is the mass loss rate $\dot M$. We will take this approach 
in this paper. The inner boundary is specified by the radius $R_\nu$ of 
the PNS, which also defines the surface of neutrino emission, 
or the neutrinosphere.
On the timescale for a mass element in the wind to receive significant
neutrino heating, the neutrino luminosities and energy spectra at
emission stay approximately fixed. Therefore, the wind can be considered
as a steady-state configuration obtained for a PNS with a fixed $R_\nu$
and a fixed mass $M_{\rm PNS}$ that emits neutrinos with fixed 
luminosities and energy spectra.

\subsection{Equations for the wind}
In this paper we ignore the effects of general relativity. To good approximation,
$v$ is nonrelativistic and the internal energy of a mass element in the wind 
is much smaller than its rest-mass energy. Based on the above considerations,
the equations describing a spherically-symmetric steady-state wind
(e.g., \citealt{qian1996nucleosynthesis}) are
\begin{subequations}
    \begin{alignat}{4}
        4\pi r^2 \rho v &=\dot{M} ,\\
        v \frac{\dd v}{\dd r} &= -\frac{1}{\rho} \frac{\dd P}{\dd r} - \frac{G M_\text{PNS}}{r^2},\\
        \frac{\dd \epsilon}{\dd r}-\frac{P}{\rho^2} \frac{\dd \rho}{\dd r} &=
        \frac{\dot{q}}{v} ,\\
        \frac{\dd Y_e}{\dd r} &=\frac{1}{v}[(\lambda_{\nu_e n}+\lambda_{e^+ n})Y_n
        -(\lambda_{\bar{\nu}_e p}+\lambda_{e^- p})Y_p],\label{eq:ye}
    \end{alignat}
\end{subequations}
where $P$ is the pressure, $G$ is the gravitational constant,
$\epsilon$ is the internal energy per unit mass,
$\dot{q}$ is the rate of net energy gain per unit mass, 
$Y_n$ ($Y_p$) is the number of free neutrons (protons) per baryon, i.e., the
number fraction of free neutrons (protons),
and $\lambda_{\nu_e n}$ ($\lambda_{e^- p}$) and 
$\lambda_{\bar{\nu}_e p}$ ($\lambda_{e^+ n}$) are the rates per target
nucleon for the reactions (inverse reactions) in 
Eqs.~(\ref{eq:nue_neutron_scattering}) and (\ref{eq:nuebar_proton_scattering}),
respectively.

The equation of state determines $P$ and $\epsilon$ as functions of
$\rho$, $T$, and the composition of the wind.
We assume that the wind is electrically neutral and composed of neutrons, 
protons, $\alpha$-particles, electrons, positrons, and photons. 
We further assume nuclear statistical equilibrium (NSE) to determine
the number fractions of the nuclear components, $Y_n$, $Y_p$, and $Y_\alpha$:
\begin{subequations}
    \begin{alignat}{3}
    Y_n+Y_p+4Y_\alpha&=1,\\
    Y_p+2Y_\alpha &=Y_e,\\
    Y_\alpha&=\frac{1}{2}Y_n^2Y_p^2\left(\frac{\rho}{m_u}\right)^3 
    \left(\frac{2\pi}{m_uT}\right)^{9/2}\exp\left(\frac{B_\alpha}{T}\right),
    \label{eq:nse}
    \end{alignat}
\end{subequations}
where $m_u$ is the atomic mass unit and
$B_\alpha\approx 28.3$~MeV is the nuclear binding energy of the 
$\alpha$-particle. We use the equation of state from \cite{timmes1999accuracy}
and \cite{timmes2000accuracy} with the above composition.
Note that in Eq.~(\ref{eq:nse}) and hereafter, the speed of light $c$, the Planck 
constant $\hbar$, and the Boltzmann constant $k$ are set to unity.

Given the inner and outer boundary conditions at the PNS and the shock,
respectively, the eigenvalue $\dot M$ can be determined by solving the
wind equations once the rates $\dot q$, $\lambda_{\nu_e n}$, $\lambda_{e^+ n}$,
$\lambda_{\bar{\nu}_e p}$, and $\lambda_{e^- p}$ are specified. In practice,
we solve for $v$, $\rho$, $T$, and $Y_e$ as functions of the radius $r$.
The corresponding $Y_n$, $Y_p$, and $Y_\alpha$ are specified by NSE and
the corresponding $P$ and $\epsilon$ are obtained from the equation of 
state.

\subsection{Rates for energy gain or loss and neutron-proton interconversion}
The rate of net energy gain per unit mass, $\dot q$, is
    \begin{alignat}{3}
        \dot q=&\ \dot q_\alpha+2\dot q_{\nu_xN}+2\dot q_{\bar\nu_xN}
        +2\dot q_{\nu_xe^\pm}+2\dot q_{\bar\nu_xe^\pm}
        +2\dot q_{\nu_x\bar\nu_x}\nonumber\\
        &+\dot q_{\nu_en}+\dot q_{\bar\nu_ep}
        +\dot q_{\nu_eN}+\dot q_{\bar\nu_eN}
        +\dot q_{\nu_ee^\pm}+\dot q_{\bar\nu_ee^\pm}
        +\dot q_{\nu_e\bar\nu_e}\label{eq:qdot}\\
        &-\dot q_{e^+n}-\dot q_{e^-p}-\dot q_{e^+e^-},\nonumber
    \end{alignat}
where
\begin{equation}
    \dot q_\alpha=v\left(\frac{\dd Y_\alpha}{\dd r}\right)\frac{B_\alpha}{m_u}
\end{equation}
corresponds to formation of $\alpha$-particles,
$\dot q_{\nu_xN}$ and $\dot q_{\nu_xe^\pm}$ to scattering of $\nu_x$ 
($\nu_\mu$ or $\nu_\tau$) on nucleons and $e^\pm$, respectively,
$\dot q_{\nu_x\bar\nu_x}$ to $\nu_x\bar\nu_x$ annihilation into $e^\pm$,
$\dot q_{\nu_en}$ ($\dot q_{e^-p}$) and $\dot q_{\bar\nu_ep}$ ($\dot q_{e^+n}$) 
to the reactions (inverse reactions) in Eqs.~(\ref{eq:nue_neutron_scattering}) 
and (\ref{eq:nuebar_proton_scattering}), respectively, and
$\dot q_{e^+e^-}$ to $e^\pm$ annihilation into neutrinos and antineutrinos of
all three active flavors. The rates denoted by other subscripts are
similar to those mentioned above. The factor of two associated with
$\nu_x$ ($\bar\nu_x$) assumes that $\nu_\mu$ and $\nu_\tau$
($\bar\nu_\mu$ and $\bar\nu_\tau$) have identical emission characteristics.

Contributions to $\dot q$, as well as the rates $\lambda_{\nu_e n}$, 
$\lambda_{e^+ n}$, $\lambda_{\bar{\nu}_e p}$, and $\lambda_{e^- p}$, can be 
separated into those that are directly affected by active-sterile neutrino
oscillations and those that are not. Those rates on the second line of
Eq.~(\ref{eq:qdot}), as well as $\lambda_{\nu_e n}$ and $\lambda_{\bar{\nu}_e p}$,
belong to the former, while those on the first and last lines
of Eq.~(\ref{eq:qdot}), as well as $\lambda_{e^+ n}$ and $\lambda_{e^- p}$,
belong to the latter. All of the rates are described in detail in 
Appendix~\ref{sec:Allrates}. Below we highlight those rates associated with
neutrino interaction processes.

We assume that all neutrinos are emitted from a neutrinosphere of radius $R_\nu$
with Fermi-Dirac spectra of zero chemical potential. For a specific neutrino
species $\nu_a$, its normalized spectrum is
\begin{equation}
    f_{\nu_a}(E)=\frac{1}{F_2(0)T_{\nu_a}^3} \frac{E^2}{\exp(E/T_{\nu_a})+1},
    \label{eq:fnu}
\end{equation}
where $T_{\nu_a}$ is the temperature parameter characteristic of $\nu_a$ emission.
In the above equation, $F_2(0)$ refers to the Fermi-Dirac integral defined as
\begin{equation}
F_k(y)=\int_0^\infty \frac{x^k}{\exp(x-y)+1}dx.
\end{equation}
At the neutrinosphere, the differential number density of $\nu_a$ per unit energy 
interval per unit solid angle is
\begin{equation}
    \left(\frac{d^2n_{\nu_a}}{dEd\Omega}\right)_{R_\nu}=
    \frac{L_{\nu_a}}{4\pi^2R_\nu^2\langle E_{\nu_a}\rangle}f_{\nu_a}(E),
\end{equation}
where $L_{\nu_a}$ and $\langle E_{\nu_a}\rangle=T_{\nu_a}F_3(0)/F_2(0)$ are
the luminosity and average energy, respectively, of $\nu_a$ at emission.

All neutrinos essentially free-stream above the neutrinosphere, but a small 
fraction can still interact, mostly with nucleons, $e^\pm$, and each other.
The angle $\theta$ between the neutrino's velocity and the radial direction 
at emission is related to the angle $\theta_r$ 
defined in the same way at radius $r$ by
\begin{equation}
\sin\theta_r=\frac{R_\nu}{r}\sin\theta.
\end{equation}
Consequently, for an interaction point at radius $r$, only those neutrinos within 
the solid angle defined by polar angles $0\leq\theta_r\leq\theta_r^{\rm max}$ are 
relevant, where $\theta_r^{\rm max}$ corresponds to a neutrino emitted tangentially 
at the neutrinosphere ($\theta=\pi/2$) and satisfies
\begin{equation}
\sin\theta_r^{\rm max}=\frac{R_\nu}{r}.
\end{equation}
Within this solid angle at radius $r$, the differential number density of $\nu_a$ 
in the absence of neutrino oscillations is
\begin{equation}
    \left(\frac{d^2n_{\nu_a}}{dEd\Omega}\right)_r=
    \left(\frac{d^2n_{\nu_a}}{dEd\Omega}\right)_{R_\nu}.
\end{equation}
In the presence of active-sterile neutrino oscillations,
the differential number density of active $\nu_a$ at radius $r$ is
\begin{equation}
    \left(\frac{d^2n_{\nu_a}}{dEd\Omega}\right)_r^{\rm osc}=
    \left(\frac{d^2n_{\nu_a}}{dEd\Omega}\right)_rP_{\nu_a}(E,r),
\end{equation}
where $P_{\nu_a}(E,r)$ is the probability for a $\nu_a$ emitted with 
energy $E$ to survive as a $\nu_a$ at radius $r$.

As an example, we give the expression of $\lambda_{\nu_en}$ at radius $r$ 
in the presence of active-sterile neutrino oscillations:
\begin{subequations}
\begin{alignat}{2}
    \lambda_{\nu_en}&=\int\left(\frac{d^2n_{\nu_e}}{dEd\Omega}\right)_r^{\rm osc}
    \sigma_{\nu_en}(E)dEd\Omega_r\\
    &=\frac{G_F^2 |V_{ud}|^2 (1+3 g_A^2)}{2\pi^2} 
    \frac{L_{\nu_e} D(r)}{R^2_\nu \langle E_{\nu_e} \rangle}
    \int_0^\infty(E+\Delta)^2\left(1-\frac{W_{\nu_e}E}{m_N}\right)f_{\nu_e}(E)P_{\nu_e}(E,r)dE,
    \label{eq:lambdanuen}
\end{alignat}
\end{subequations}
where
\begin{equation}
    	\sigma_{\nu_en}(E)=\frac{G_F^2 |V_{ud}|^2 (1+3 g_A^2)}{\pi} 
        (E+\Delta)^2\left(1-\frac{W_{\nu_e}E}{m_N}\right)
        \label{eq:sigmanuen}
\end{equation}
is the cross section for $\nu_e$ absorption on neutrons and
\begin{equation}
D(r)=\frac{1}{2\pi}\int d\Omega_r=\int_0^{\theta_r^{\rm max}}\sin\theta_rd\theta_r
=1-\sqrt{1-\frac{R_\nu^2}{r^2}}.
\end{equation}
In Eq.~(\ref{eq:sigmanuen}), $\Delta\equiv m_n-m_p$ is the neutron-proton mass 
difference, $m_N=(m_n+m_p)/2$ is the average nucleon mass, the term associated with
\begin{equation}
	W_{\nu_e} = \frac{2[1 + 5 g_A^2 - 2 g_A (1 + f_2)]}{1 + 3 g_A^2}
\end{equation}
is the correction due to weak magnetism and nucleon recoil, and
$G_F$, $V_{ud}$, $g_A$, and $f_2$ are the standard constants describing the weak 
interaction of concern.

\subsection{Active-sterile neutrino oscillations}
We only consider $\nu_e$-$\nu_s$ and $\bar\nu_e$-$\bar\nu_s$ oscillations in this 
paper. We assume that neutrinos emitted with energy $E$ but from different points 
on the neutrinosphere have the same flavor evolution as those traveling on the 
radial trajectory. For a radially-propagating $\nu_e$ emitted with energy
$E$, the evolution of its wave function
\begin{equation}
\psi_{\nu_e}(E,r)=\left(\begin{array}{c}
a_{\nu_e}(E,r)\\
a_{\nu_s}(E,r)
\end{array}\right)
\end{equation}
is governed by
\begin{equation}
i\dfrac{d}{dr}\psi_{\nu_e}(E,r)=\mathbf{H}_{\nu_e}(E,r)\psi_{\nu_e}(E,r),
\label{eq:dpsidr}
\end{equation}
where $a_{\nu_e}$ ($a_{\nu_s}$) is the amplitude for being a $\nu_e$ ($\nu_s$),
and
\begin{equation}
\mathbf{H}_{\nu_e}(E,r)=\mathbf{H}_\mathrm{vac}(E)+\mathbf{H}_\mathrm{mat}(r)+
\mathbf{H}_{\nu\nu}(r)
\end{equation}
is the propagation Hamiltonian. The three contributions to this Hamiltonian are
\begin{equation}
\mathbf{H}_\mathrm{vac}(E)=\frac{\delta m^2}{4E}
\left(\begin{array}{rc}
-\cos2\theta_V&\sin2\theta_V\\
\sin2\theta_V&\cos2\theta_V\end{array}\right),
\end{equation}
where $\delta m^2$ is the vacuum mass-squared difference and $\theta_V$ is the
vacuum mixing angle,
\begin{equation}   
\mathbf{H}_\text{mat}(r)=\frac{V_\text{mat}(r)}{2}
\left(\begin{array}{cc}
1&0\\
0&-1\end{array}\right),
    \label{eq:matterTerm}
\end{equation}
where
\begin{equation}
V_\text{mat}(r)=\frac{\sqrt{2}G_F}{2m_u}\rho(r)[3Y_e(r)-1]
\label{eq:Vmat}
\end{equation}
corresponds to neutrino forward scattering on $e^\pm$, neutrons, and 
protons (\citealt{wolfenstein1978neutrino,mikheev1985resonance}, see also, e.g.,
\citealt{sigl1993general}), and
\begin{equation}
\mathbf{H}_{\nu\nu}(r)=\frac{V_\nu(r)}{2}
\left(\begin{array}{cc}
1&0\\
0&-1\end{array}\right),
\label{eq:nuTerm}
\end{equation}
where
\begin{equation}
V_{\nu}(r)
=\frac{\sqrt{2} G_F D(r)^2}{2\pi R_\nu^2}
\int\left[\frac{L_{\nu_e}}{\langle E_{\nu_e} \rangle} f_{\nu_e}(E') P_{\nu_e}(E',r) -  
\frac{L_{\bar{\nu}_e}}{\langle E_{\bar{\nu}_e} \rangle} f_{\bar{\nu}_e}(E') 
{P}_{\bar{\nu}_e}(E',r) \right]dE'
\label{eq:Vnu}
\end{equation}
corresponds to neutrino forward scattering on other neutrinos 
(e.g., \citealt{fuller1987resonant,sigl1993general}).
In Eq.~(\ref{eq:Vnu}), $P_{\bar{\nu}_e}(E,r)$ is calculated from the amplitude
$a_{\bar{\nu}_e}(E,r)$, which in turn is determined along with $a_{\bar{\nu}_s}(E,r)$
by the same flavor evolution equation as Eq.~(\ref{eq:dpsidr}), except with
the Hamiltonian 
$\mathbf{H}_{\bar\nu_e}(E,r)=\mathbf{H}_\mathrm{vac}(E)-\mathbf{H}_\mathrm{mat}(r)
-\mathbf{H}_{\nu\nu}(r)$. 
Because we only consider $\nu_e$-$\nu_s$ and $\bar\nu_e$-$\bar\nu_s$ oscillations,
there is no flavor evolution for $\nu_x$ or $\bar\nu_x$. We further assume that
$\nu_x$ and $\bar\nu_x$ have the same emission characteristics so that their
contributions to $\mathbf{H}_{\nu\nu}(r)$ cancel.

\subsection{Parameters and procedures}
A number of experiments indicate the existence of a $\nu_s$ ($\bar\nu_s$) that mixes 
with $\nu_e$ ($\bar\nu_e$) for a range of plausible vacuum mixing parameters. As our
major goal is to include active-sterile neutrino oscillations in a self-consistent
treatment of the wind and to demonstrate their effects, we follow \cite{wu2014impact}
and consider three sets of $\delta m^2$ and $\sin^22\theta_V$ that are listed in 
Table~\ref{tab:NDWnumericalOscillationParameters} and referred to as Cases A, B, and C,
respectively. For comparison, the case without oscillations is referred to as Case D.
We plan to survey a broad range of mixing parameters in a separate paper.

\begin{table}[!hbt]
    \caption{Cases of active-sterile neutrino oscillations}
    \label{tab:NDWnumericalOscillationParameters}
    \begin{center}
    \begin{tabular}{ccc}\hline\hline
          & $\delta m^2$ (eV$^2$) & $\sin^2 2\theta_V$ \\\hline
        A & 1.75 & 0.10 \\\hline
        B & 1.0  & 0.06 \\\hline
        C & 0.4  & 0.04 \\\hline
        D & no  & oscillations \\\hline
    \end{tabular}
    \end{center}
\end{table}

To solve the wind equations, we need to specify properties of the PNS, especially
the characteristics of its neutrino emission, the inner boundary conditions at the
neutrinosphere, and the outer boundary conditions at the CCSN shock. We use the 
spherical CCSN model of \cite{martinez2014supernova} as a guide and consider
three snapshots at time post core bounce $t_{\rm pb}=1$, 2, and 5~s, respectively.
The PNS mass is taken to be fixed at $M_{\rm PNS}=1.282M_\odot$. The neutrinosphere
radius $R_\nu$ for each snapshot is given in Table~\ref{tab:NDWnumericalDynamicParameters}
along with the luminosities and average energies of $\nu_e$, $\bar\nu_e$, and $\nu_x$
at emission. As noted above, the emission characteristics of $\bar\nu_x$ are taken 
to be identical to those of $\nu_x$. We assume that the shock launched by the CCSN
explosion is unaffected by active-sterile neutrino oscillations. For the epoch of
interest, the region at $r\gtrsim 10^3$~km enclosed by the shock has approximately 
the same temperature, which slowly decreases as the shock moves outward. We take the 
temperature at $r=10^3$~km~$\equiv R_b$ as the effective outer boundary condition set up by the 
CCSN shock. The values of $T(R_b)$ are given in 
Table~\ref{tab:NDWnumericalDynamicParameters}
along with the corresponding temperatures $T(R_\nu)$ at the neutrinosphere.
In addition to $T(R_\nu)$, the other inner boundary conditions are values
of $\rho(R_\nu)$ and $Y_e(R_\nu)$, which are determined by assuming
$\dot{q}=0$ and $dY_e/dr=0$ at the neutrinosphere. Note that for the vacuum mixing 
parameters of interest, $\nu_e$-$\nu_s$ and $\bar\nu_e$-$\bar\nu_s$ oscillations
can be ignored at the neutrinosphere and $\rho(R_\nu)$ and $Y_e(R_\nu)$ correspond 
to conditions in the absence of neutrino oscillations. Accordingly, we assume that 
neutrinos and antineutrinos are in their pure active flavor states at the neutrinosphere.

\begin{table}[!hbt]
    \caption{Parameters at the inner and outer boundaries. Neutrino luminosities are given
    in units of $10^{51}$~erg~s$^{-1}$. Average neutrino energies and temperatures are given in
    units of MeV.}
    \label{tab:NDWnumericalDynamicParameters}
    \begin{center}
    \begin{tabular}{cccccccccc}\hline\hline
        $t_\text{pb}$ (s) & $R_\nu$ (km) & $L_{\nu_e}$ & 
        $L_{\bar{\nu}_e}$ & $L_{\nu_x}$ & 
        $\langle E_{\nu_e} \rangle$ & $\langle E_{\bar{\nu}_e} \rangle$ & 
        $\langle E_{\nu_x} \rangle$ &
        $T(R_\nu)$ & $T(R_b)$ \\\hline
        1 & 22.25 & 2.95 & 3.68 & 3.83 & 8.80 & 12.49 & 12.34 & 2.79 & 0.17 \\\hline
        2 & 18.07 & 1.67 & 2.01 & 2.58 & 8.43 & 11.90 & 11.70 & 2.68 & 0.14\\\hline
        5 & 14.50 & 0.82 & 0.83 & 1.42 & 7.79 & 10.17 & 10.47 & 2.47 & 0.09\\\hline
    \end{tabular}
    \end{center}
\end{table}

For a specific set of $\delta m^2$ and $\sin^22\theta_V$ and a specific set of
boundary conditions, we use the following procedure to determine $v(r)$, $\rho(r)$,
$T(r)$, and $Y_e(r)$ for the wind: (1) calculate $\rho(R_\nu)$ and $Y_e(R_\nu)$ for 
the corresponding $T(R_\nu)$ using $\dot q(R_\nu)=0$ and $(dY_e/dr)_{R_\nu}=0$, 
(2) pick an estimated value of $\dot{M}$ and calculate $v(R_\nu)$,
(3) starting from $r=R_\nu$, integrate the wind equations 
along with the equations of $\nu_e$ and $\bar\nu_e$ 
flavor evolution up to $r=R_b$, (4) repeat steps 2 and 3 until the value of
$T(R_b)$ given by the wind solution satisfies the outer boundary condition.
The corresponding $\dot M$ is the eigenvalue of the mass loss rate for the wind.

\section{Results on neutrino-driven winds} \label{sec:results}
Below we present the results on the wind for three sets of vacuum mixing parameters 
(cases A, B, and C in Table~\ref{tab:NDWnumericalOscillationParameters}) and
for three snapshots of the CCSN evolution ($t_{\rm pb}=1$, 2, and 5~s). We also 
compare these results with those in the absence of neutrino oscillations.

\subsection{Mass loss rates and profiles of $v$, $\rho$, and $T$}
\begin{figure}[t]
\centering
    \includegraphics[width=0.99\textwidth]{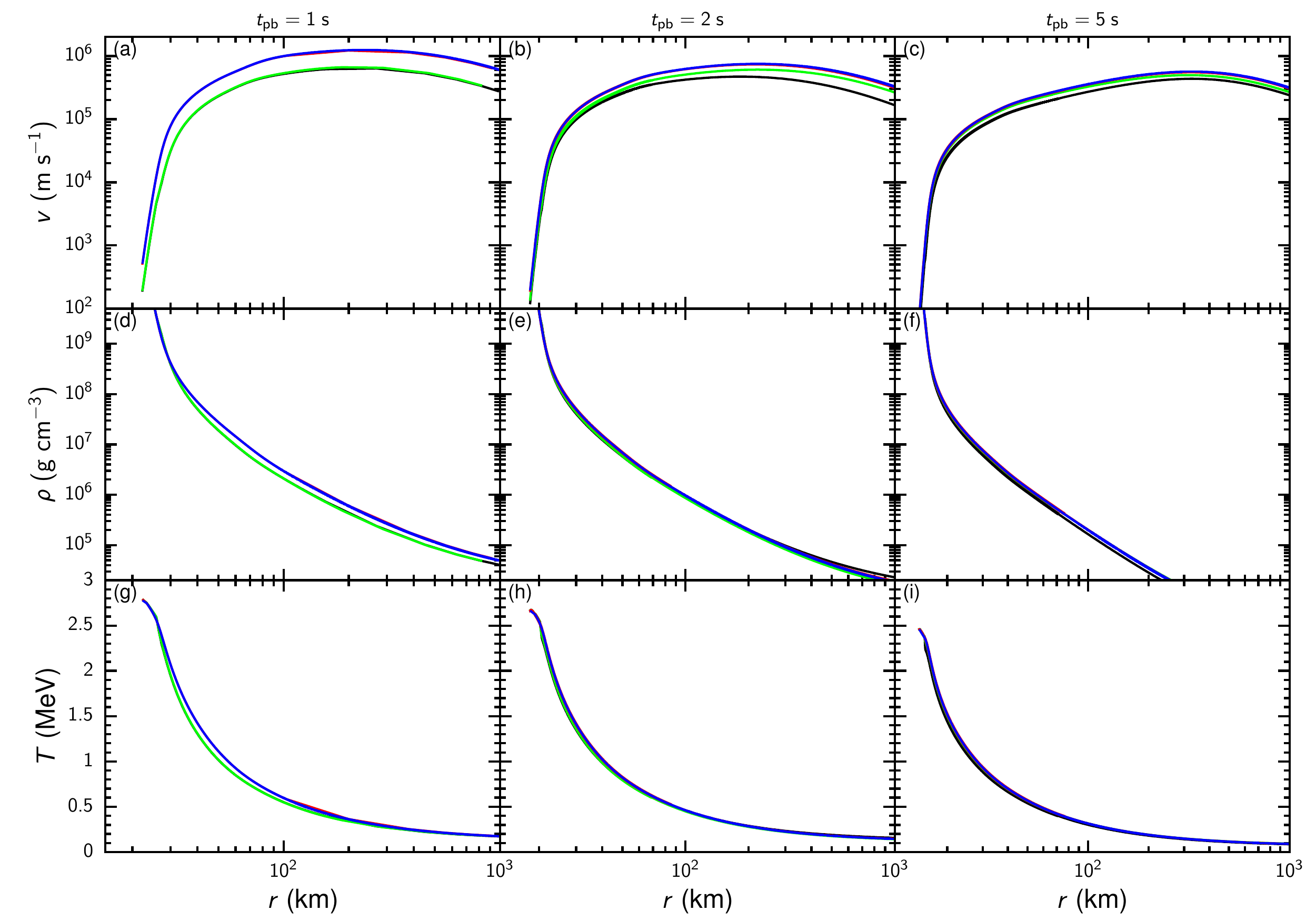}
\caption{\label{fig:dynamicAll} Wind profiles of $v$ (upper row), $\rho$ (central row), 
and $T$ (bottom row) as functions of radius $r$ at $t_{\rm pb}=1$~s 
(left column), 2~s (central column), and 5~s (right column) for 
cases A (black), B (green), C (red), and D (blue).
Note that red and blue curves are indistinguishable.
}
\end{figure}

The results on the wind mass loss rate $\dot M$ for each case of interest are
given in Table~\ref{tab:NDWnumericalMdot}, and those on the profiles of $v$, 
$\rho$, and $T$ are shown in Fig.~\ref{fig:dynamicAll}. For all three snapshots,
the wind solutions for the vacuum mixing parameters of case C are very close to those
in the absence of neutrino oscillations (case D), whereas the solutions for
case A show the largest deviations. The solutions for case B are similar to those
for case A for $t_{\rm pb}=1$~s, but become similar to case C for $t_{\rm pb}=5$~s.
In addition, the effects of oscillations on $\dot M$ and the profile of $v$ are much 
more significant than those on the profiles of $\rho$ and $T$. As case A corresponds
to the largest effects of oscillations, we focus on this case in this subsection.

\begin{table}[!hbt]
    \caption{Mass loss rates in units of $M_\odot$~s$^{-1}$. The letters refer to the sets 
    of vacuum mixing parameters in Table~\ref{tab:NDWnumericalOscillationParameters} and $X(Y)$ 
    denotes $X\times10^Y$.}
    \label{tab:NDWnumericalMdot}
    \begin{center}
    \begin{tabular}{ccccc}\hline\hline
        $t_{\rm pb}$ (s)    & A & B & C & D \\\hline
        1 & $6.95(-5)$ & $7.10(-5)$ & $1.86(-4)$ & $1.87(-4)$ \\\hline
        2 & $2.40(-5)$ & $2.76(-5)$ & $3.80(-5)$ & $3.81(-5)$ \\\hline
        5 & $2.82(-6)$ & $4.16(-6)$ & $4.50(-6)$ & $4.51(-6)$ \\\hline
    \end{tabular}
    \end{center}
\end{table}

\begin{figure}[!ht]
\centering
    \includegraphics[width=0.32\textwidth]{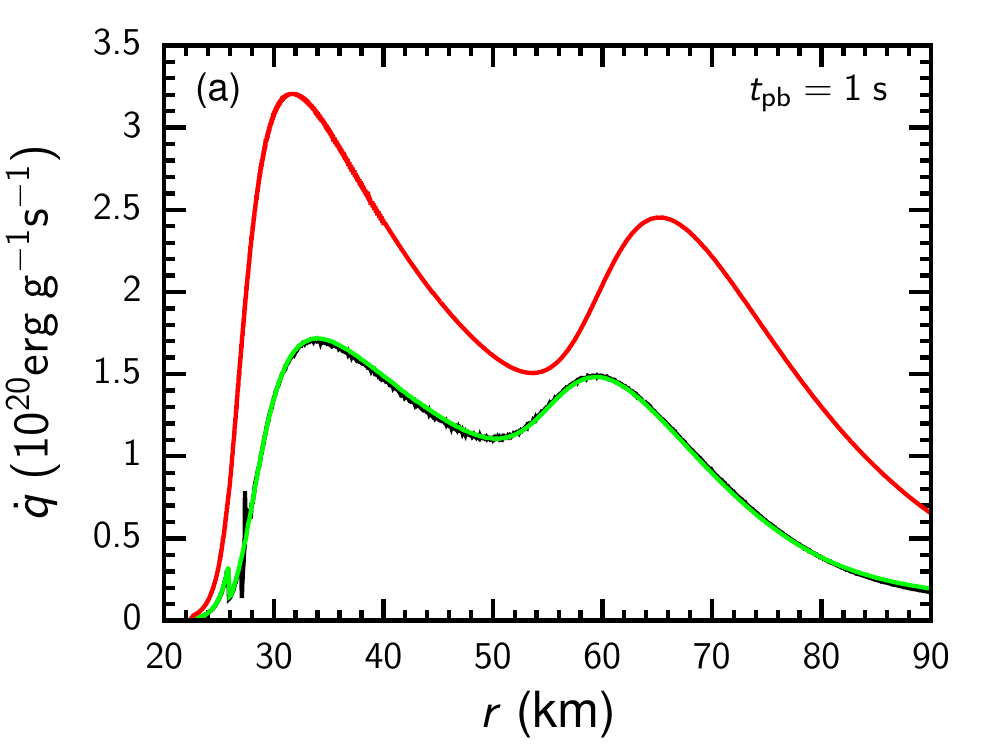}
    \includegraphics[width=0.32\textwidth]{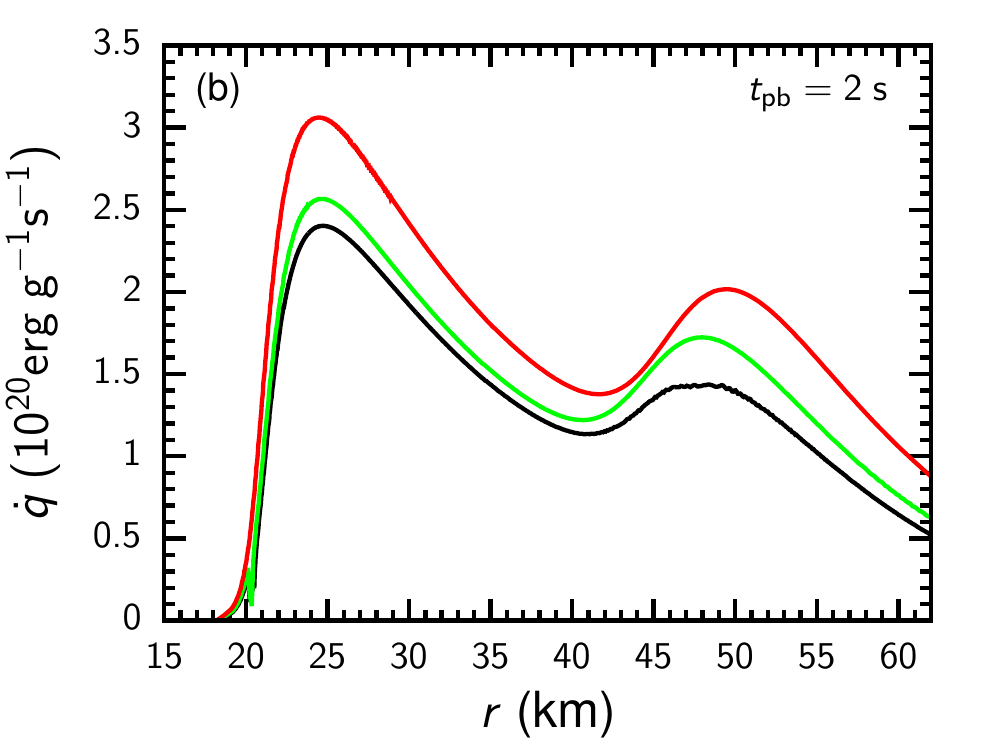}
    \includegraphics[width=0.32\textwidth]{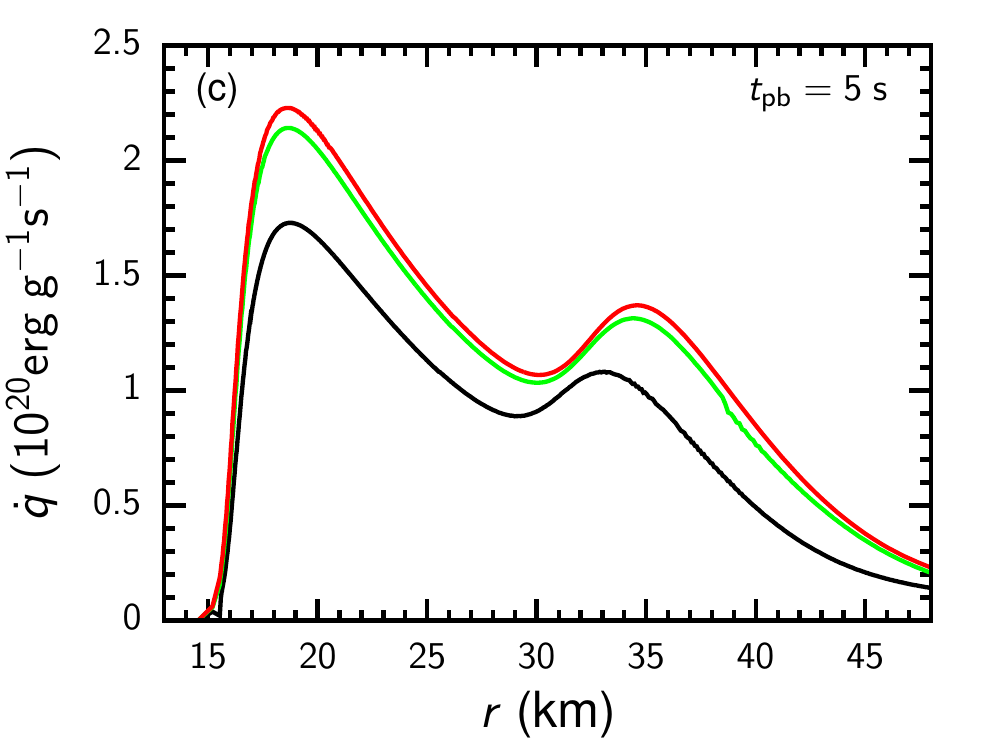}
    \includegraphics[width=0.32\textwidth]{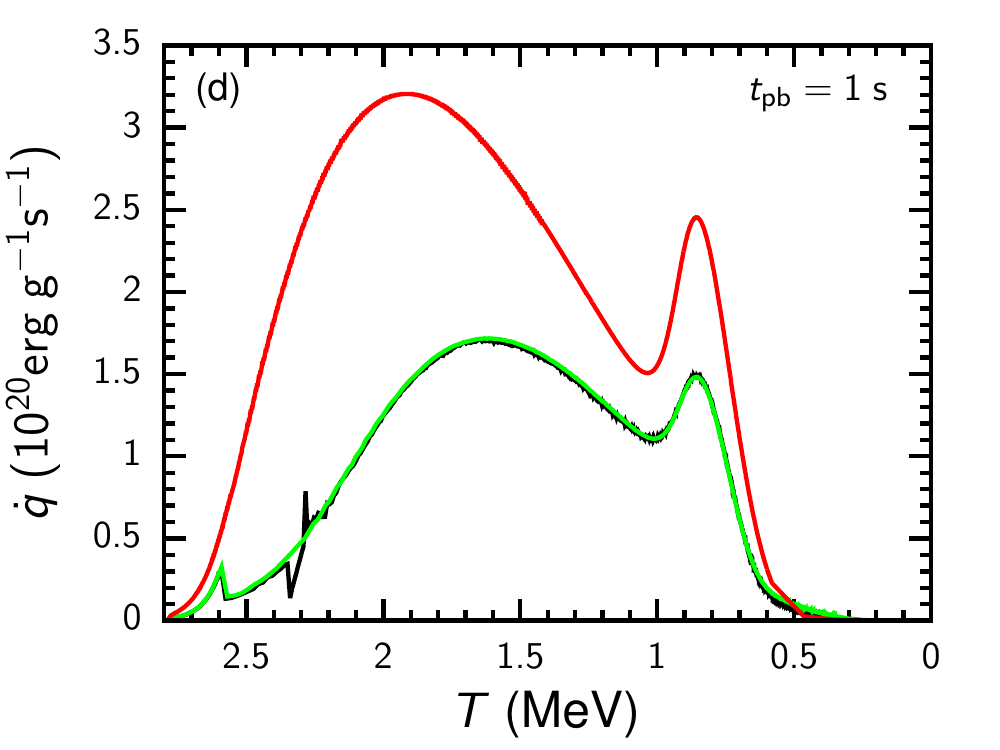}
    \includegraphics[width=0.32\textwidth]{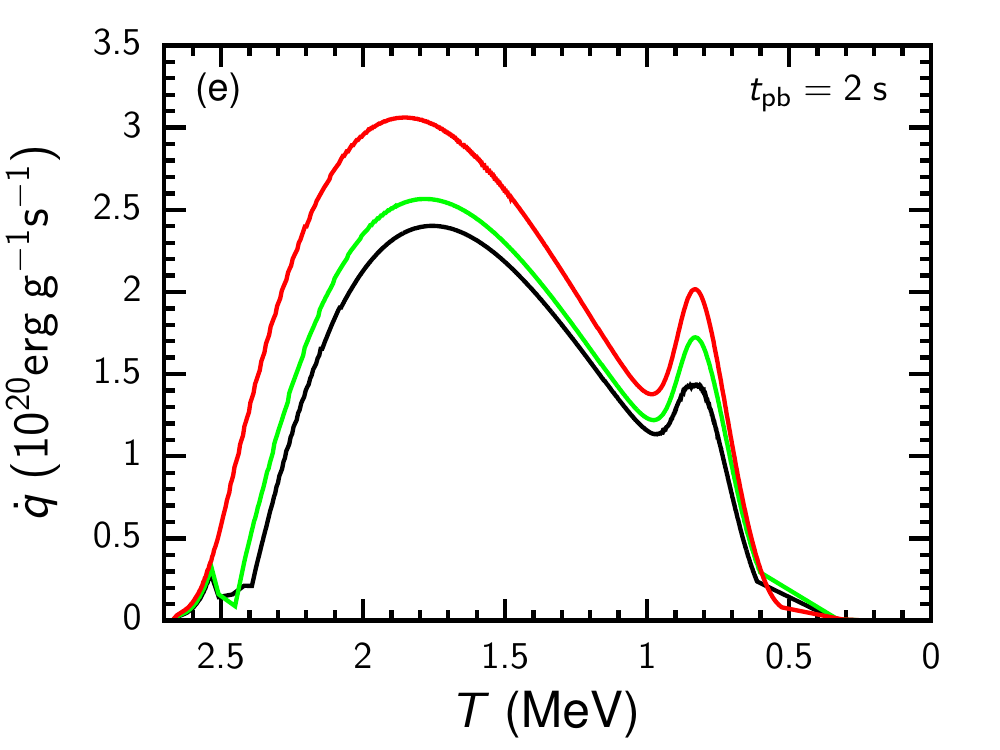}
    \includegraphics[width=0.32\textwidth]{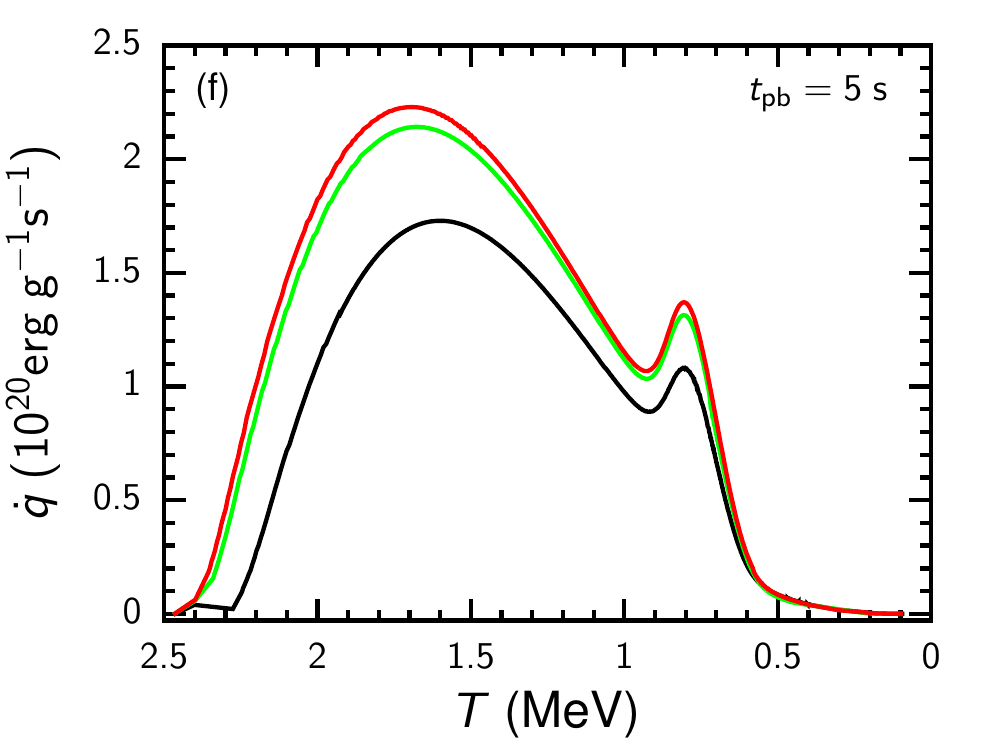}
    \includegraphics[width=0.32\textwidth]{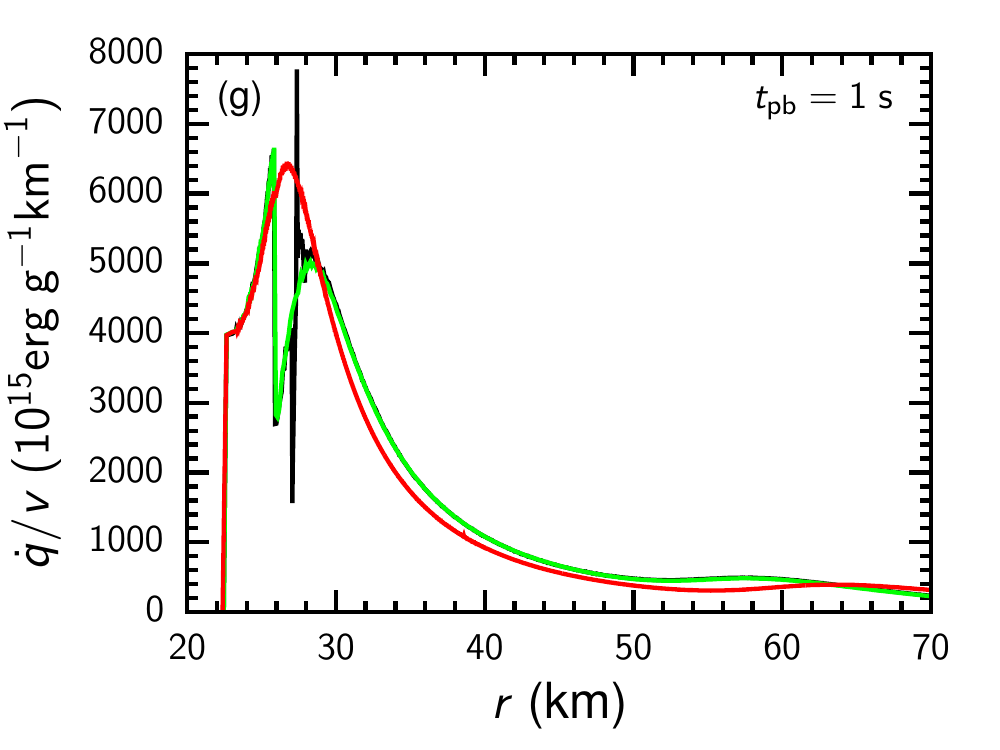}
    \includegraphics[width=0.32\textwidth]{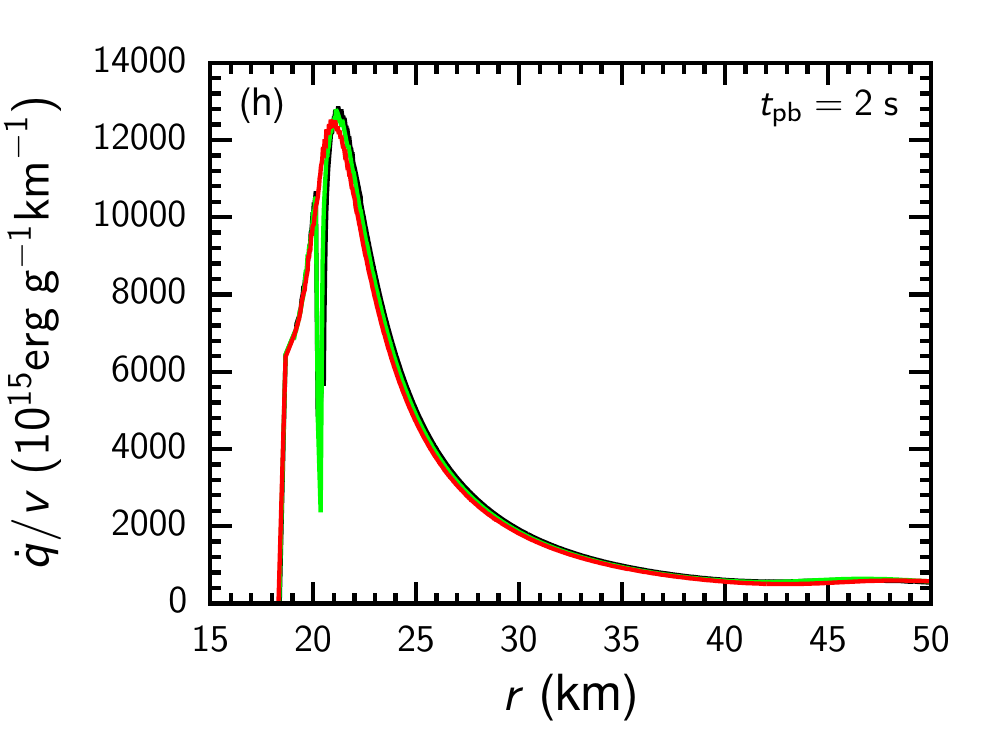}
    \includegraphics[width=0.32\textwidth]{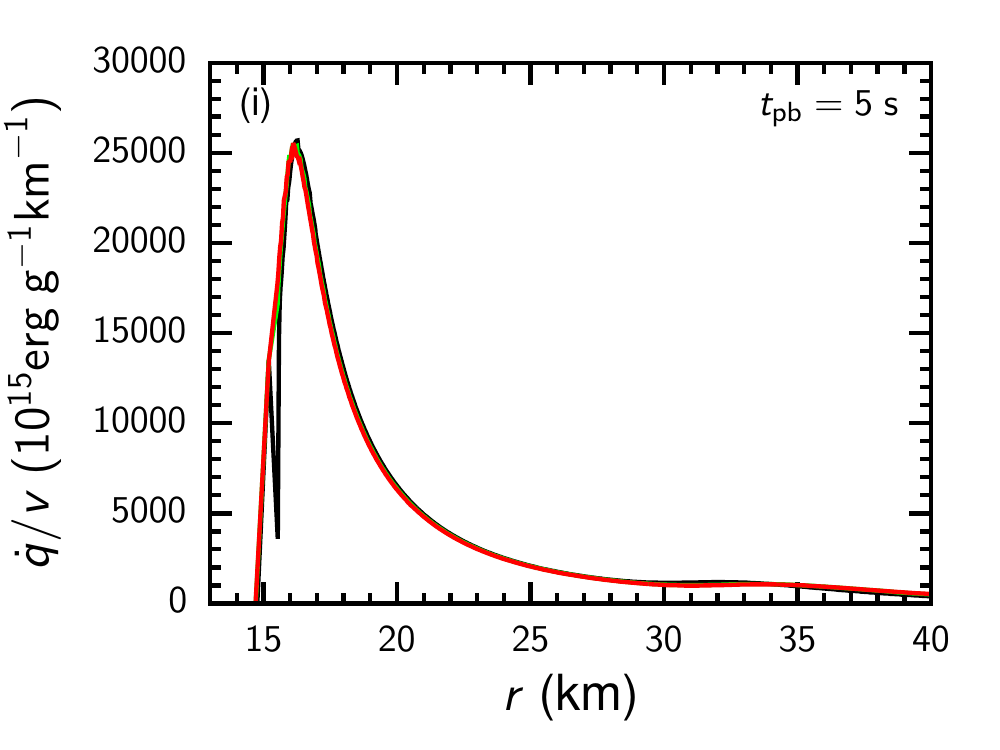}
    \includegraphics[width=0.32\textwidth]{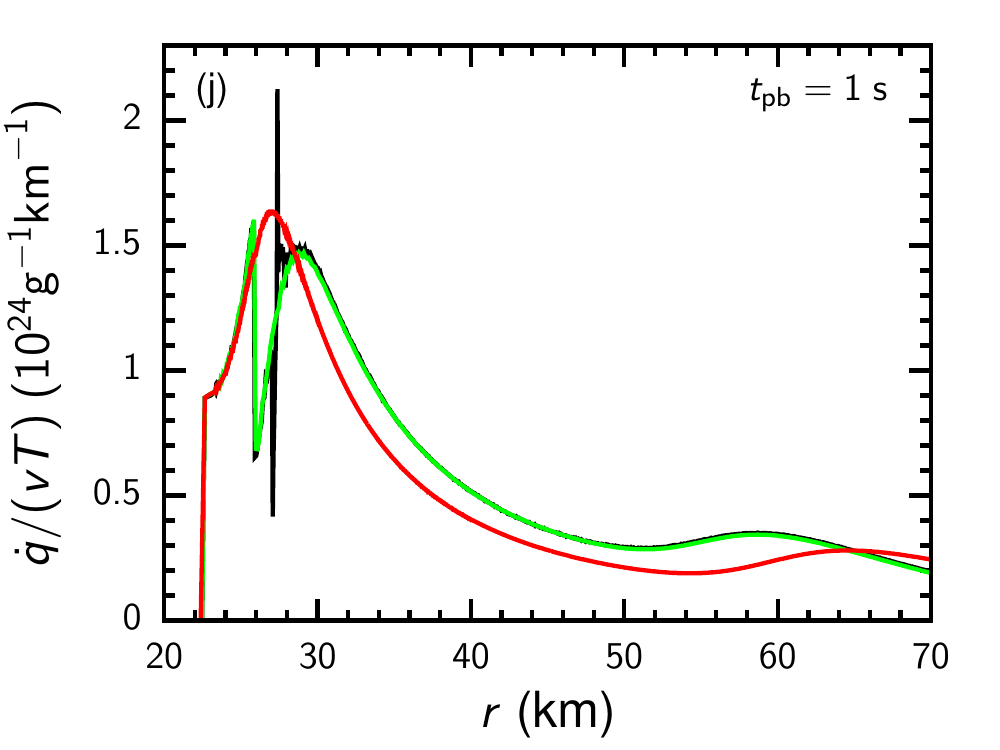}
    \includegraphics[width=0.32\textwidth]{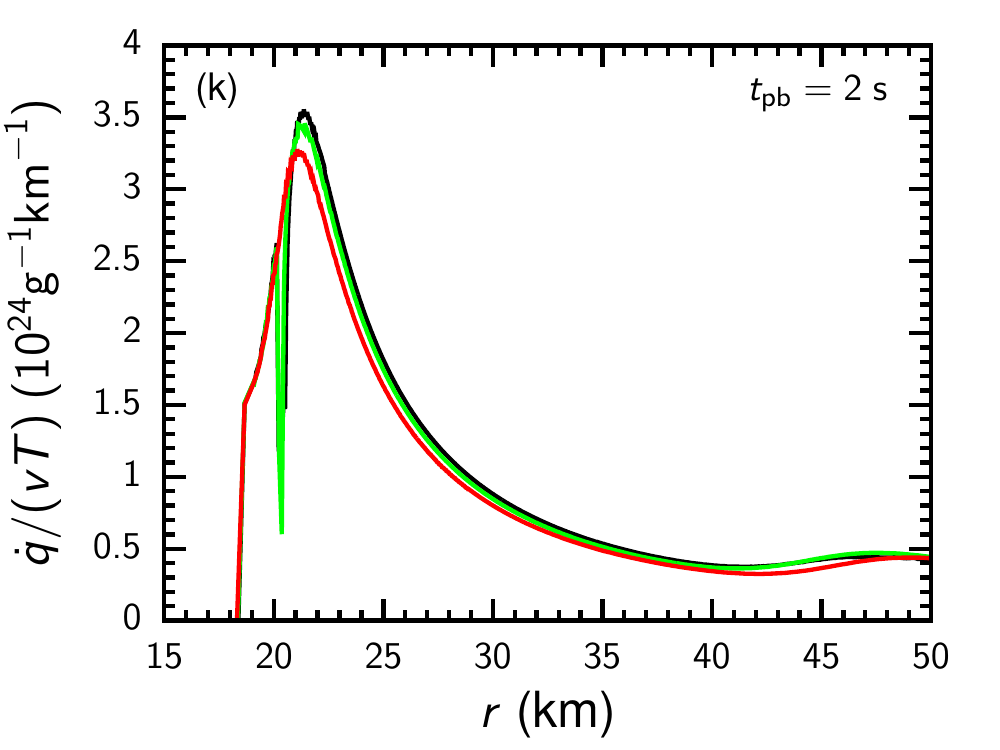}
    \includegraphics[width=0.32\textwidth]{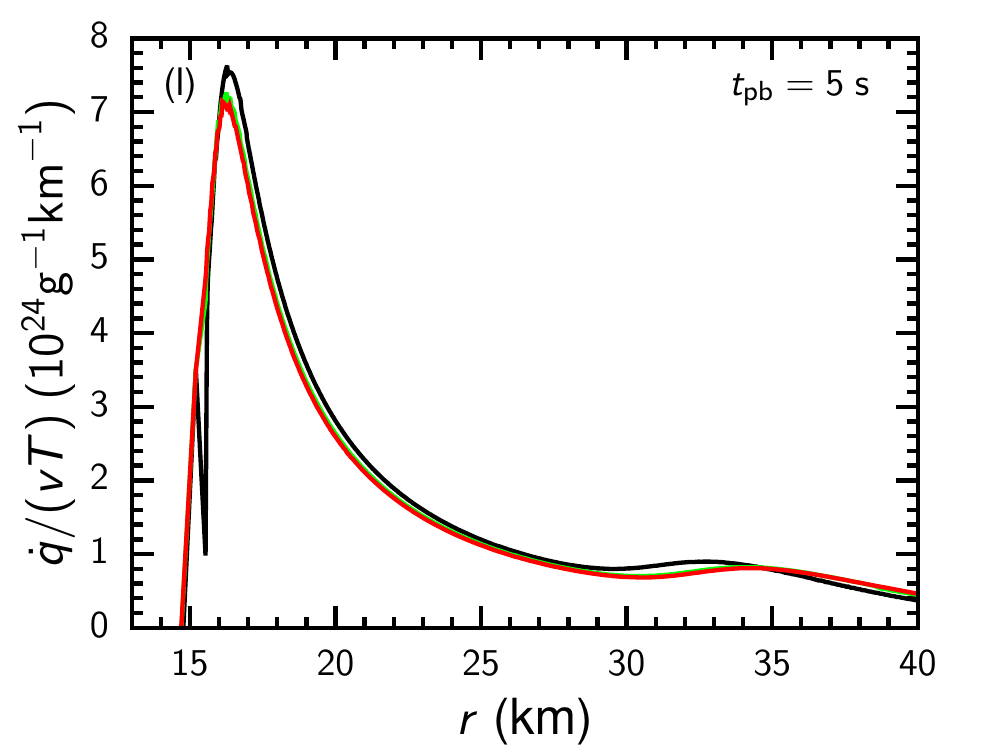}
    \caption{\label{fig:heating} Net heating rates $\dot{q}$ as functions of 
radius $r$ (upper row) and temperature $T$ (second row), as well as $\dot{q}/v$ (third row) 
and $\dot{q}/(vT)$ (bottom row) as functions of $r$
at $t_{\rm pb}=1$~s (left column), 2~s (central column), and 5~s (right column) 
for cases A (black), B (green), and C (red). 
}
\label{fig:qdot}
\end{figure}

The rate of net energy gain per unit mass $\dot q$ is crucial to the understanding
of the wind. This quantity is shown as functions of $r$ and $T$, respectively, in 
Figs.~\ref{fig:qdot}(a)-(f). Our inner boundary conditions require $\dot q(R_\nu)=0$. 
Because the rate of cooling is sensitive to $T$, it decreases more rapidly than the rate 
of heating by neutrinos at $r>R_\nu$. The resulting net energy gain causes $\dot q$ to 
increase sharply in the immediate neighborhood of $r=R_\nu$. However, the decrease of 
the neutrino heating rate due to the diminishing neutrino fluxes 
and any $\nu_e$-$\nu_s$ and $\bar\nu_e$-$\bar\nu_s$ oscillations takes effect at larger
$r$. The above two competing factors cause $\dot q$ to peak near the neutrinosphere. 
At even larger $r$ corresponding to $T<1$~MeV, formation of $\alpha$-particles 
becomes important and the associated energy release results in the other peak of $\dot q$.
At $T\gtrsim 1$~MeV, neutrino heating by the reactions in 
Eqs.~(\ref{eq:nue_neutron_scattering}) and (\ref{eq:nuebar_proton_scattering}) 
makes the dominant contribution to $\dot q$. As the rate of heating by these reactions is
substantially reduced by $\nu_e$-$\nu_s$ and $\bar\nu_e$-$\bar\nu_s$ oscillations in case A,
the $\dot q$ in this case is significantly less than that in case D without oscillations.
In order to drive the wind, the net energy gain must overcome 
the gravitational potential of the PNS \citep{qian1996nucleosynthesis}. So the mass loss
rate of the wind can be estimated as
\begin{equation}
\dot M\sim\frac{R_\nu}{GM_{\rm PNS}}\int_{R_\nu}^{R_b}4\pi r^2\rho\dot q dr.
\label{eq:heat1}
\end{equation}
The reduction of $\dot q$ by oscillations in case A leads to
the reduction of $\dot M$ by a factor of $\approx 2.7$, 1.6, and 1.6 for $t_{\rm pb}=1$, 2, 
and 5~s, respectively, relative to case D (see Table~\ref{tab:NDWnumericalMdot}).

Using $4\pi r^2\rho\dot q=\dot M\dot q/v$, we can rewrite Eq.~(\ref{eq:heat1}) as
\begin{equation}
\int_{R_\nu}^{R_b}\frac{\dot q}{v}dr\sim\frac{GM_{\rm PNS}}{R_\nu}.
\label{eq:heat2}
\end{equation}
Because the right-hand side of the above equation is fixed for both cases with 
and without oscillations, the reduction of $\dot q$ by oscillations
in case A is expected to be accompanied by the reduction of $v$ relative to case D
(see Figs.~\ref{fig:dynamicAll}(a)-(c)). With the reduced neutrino heating rate,
the wind material in case A must move more slowly so that there is more time for 
the material to gain the energy required to overcome the gravitational potential of 
the PNS. We show $\dot q/v$ as a function of $r$ in Figs.~\ref{fig:qdot}(g)-(i). It can be
seen that except for the region near the neutrinosphere where oscillations give rise
to complicated behaviors in case A, $\dot q/v$ is very similar for cases A and D.

\begin{figure}[!ht]
\centering
    \includegraphics[width=0.32\textwidth]{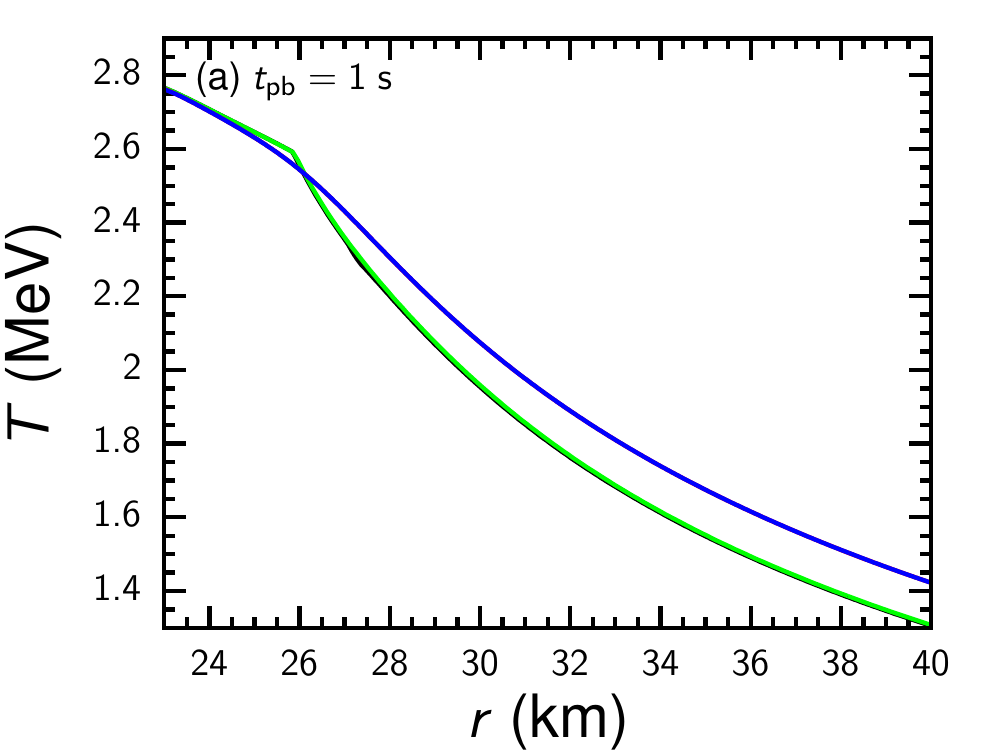}
    \includegraphics[width=0.32\textwidth]{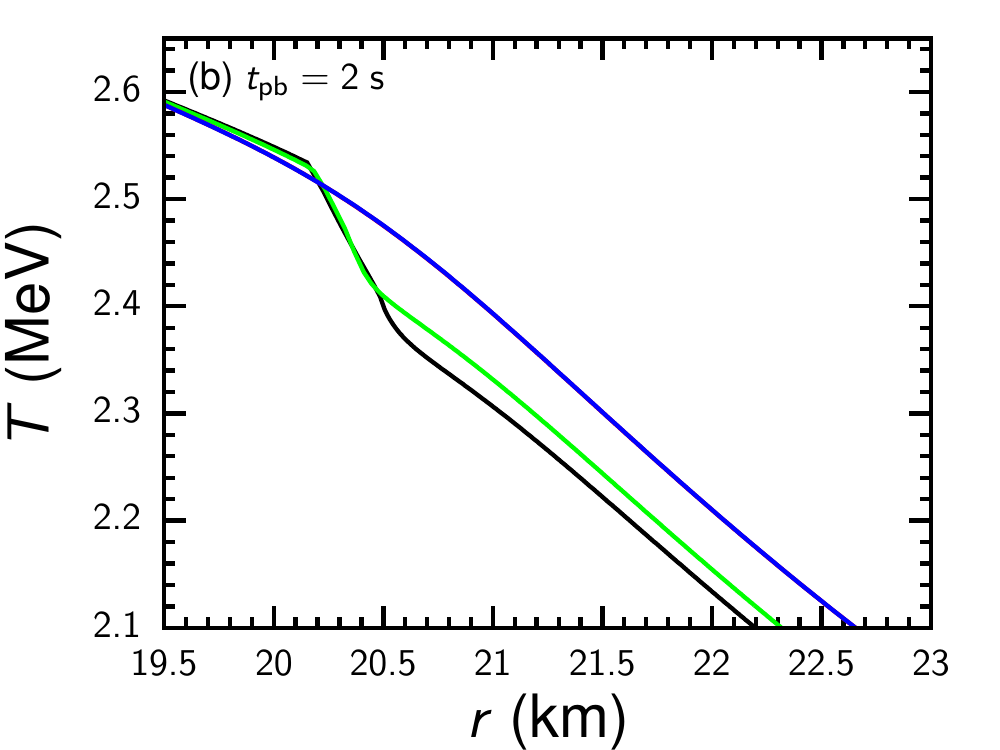}
    \includegraphics[width=0.32\textwidth]{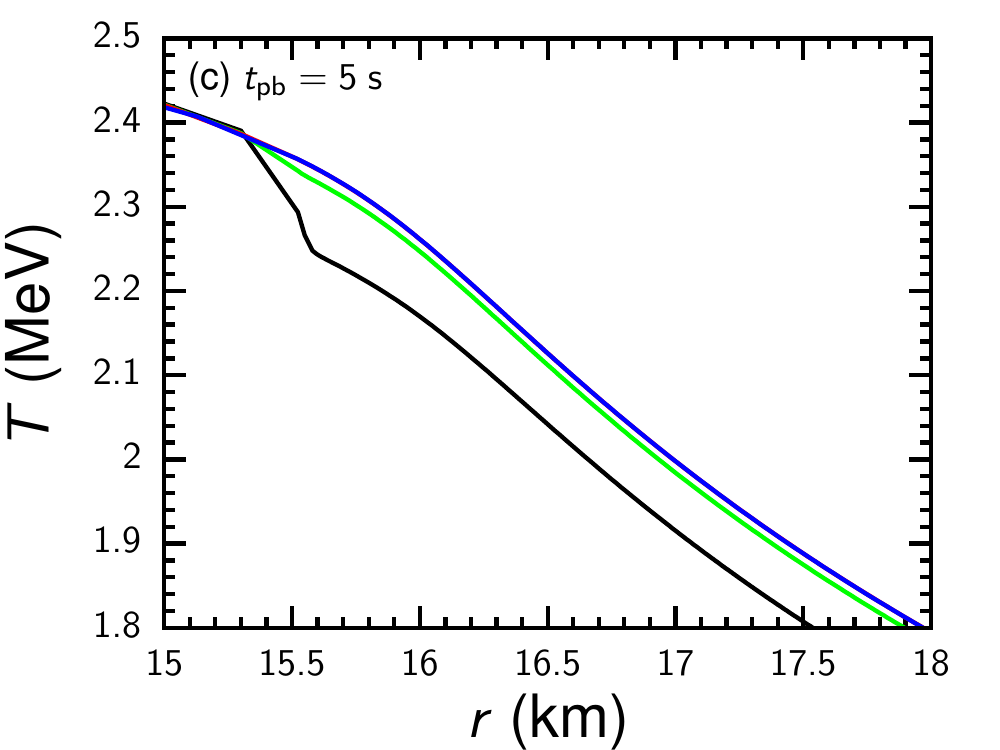}
\caption{\label{fig:dynamic2s} Wind temperature $T$ near the onset of active-sterile
neutrino oscillations as a function of radius $r$
at $t_{\rm pb}=1$~s (a), 2~s (b), and 5~s (c) for cases 
A (black), B (green), C (red), and D (blue). 
Note that red and blue curves are indistinguishable.
}
\end{figure}

\begin{figure}[t]
\centering
    \includegraphics[width=0.99\textwidth]{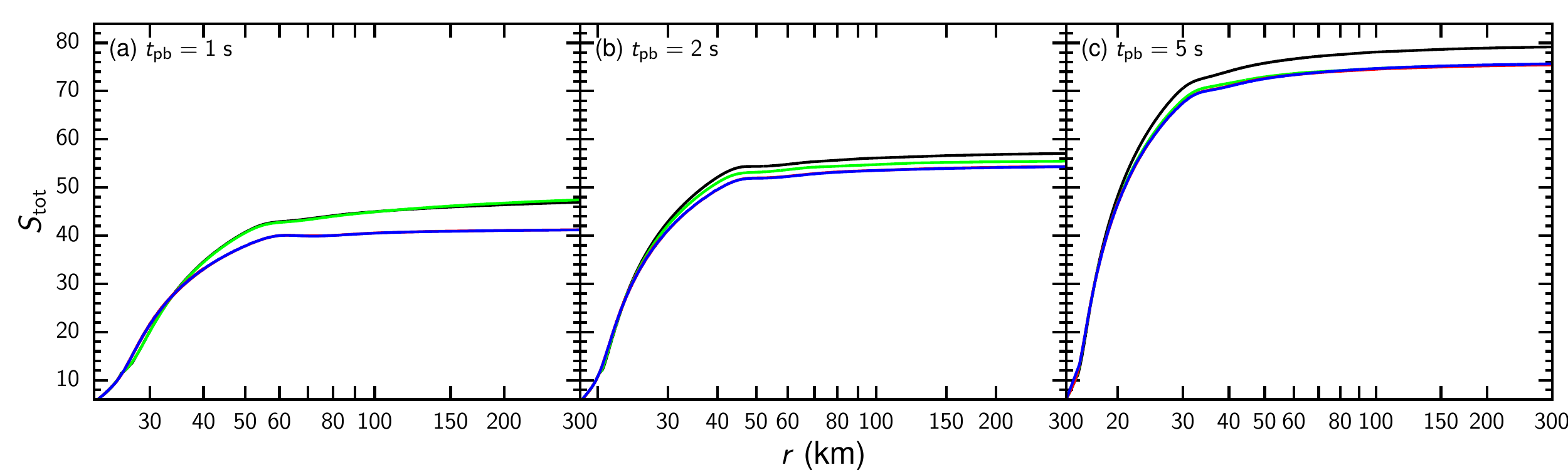}
\caption{\label{fig:EntropyAll} Total entropy $S_{\rm tot}$ as a function of radius $r$
at $t_{\rm pb}=1$~s (a), 2~s (b), and 5~s (c) for cases 
A (black), B (green), C (red), and D (blue). 
Note that red and blue curves are indistinguishable.
}
\end{figure}

Compared to $v$, the differences in the profile of $\rho$ or $T$ between cases A and D
are much smaller (see Fig.~\ref{fig:dynamicAll}). This result is expected for $T$ because 
$T(R_\nu)$ and $T(R_b)$ are fixed at the inner and outer boundaries for all cases. 
Indeed, Fig.~\ref{fig:dynamic2s} shows that relative to case D,
the reduction in the neutrino heating rate by oscillations in case A causes a decrease 
in $T$ by at most $\sim 6\%$, 3\%, and 4\% for $t_{\rm pb}=1$, 2, and 5~s, respectively.
This initial decrease is compensated as the slower wind material in case A gains energy 
at the reduced neutrino heating rate over a longer time than case D so that the same 
$T(R_b)$ is obtained. The similar profiles of $T$ and $\dot q/v$
for cases A and D result in similar profiles of $\dot q/(vT)$ (see Figs.~\ref{fig:qdot}(j)-(l)).
Because the total entropy per baryon $S_{\rm tot}$ as a function of $r$ 
can be estimated as
\begin{equation}
S_{\rm tot}(r)\sim S_{\rm tot}(R_\nu)+\int_{R_\nu}^r\frac{m_u\dot q}{Tv}dr,
\end{equation}
similar profiles of $\dot q/(vT)$ lead to similar profiles of $S_{\rm tot}$ for
cases A and D. As shown in Fig.~\ref{fig:EntropyAll}, relative to case D,
the largest deviation of $S_{\rm tot}$ for case A is an increase by $\sim 16\%$, 5\%, and 5\%
for $t_{\rm pb}=1$, 2, and 5~s, respectively. Because $S_{\rm tot}(r)$ mainly depends on
$T$ and $\rho$, being approximately $\propto T^3/\rho$ \citep{qian1996nucleosynthesis}, 
similar profiles of $S_{\rm tot}$ and $T$ mean similar profiles of $\rho$ as well.
As $\dot M=4\pi r^2\rho v$, the reduction of $\dot M$ by oscillations in case A mostly
translates into the reduction of $v$ relative to case D, which is in agreement with
the discussion in the preceding two paragraphs.

\subsection{Active-sterile neutrino oscillations and effects on the $Y_e$ profile}
In addition to the effects on the mass loss rates and profiles of $v$, $\rho$, and $T$
discussed above, active-sterile neutrino oscillations can change the evolution of $Y_e$, 
which is critical to nucleosynthesis. Further, the intricate feedback between the evolution
of $Y_e$ and survival probabilities of $\nu_e$ and $\bar\nu_e$ leads to a range of 
interesting behaviors of active-sterile neutrino oscillations, which are presented below.
For convenience, we denote the cases under discussion by the letter for 
the vacuum mixing parameters (see Table~\ref{tab:NDWnumericalOscillationParameters}) 
followed by the numerical value of the $t_{\rm pb}$ for the snapshot 
(see Table~\ref{tab:NDWnumericalDynamicParameters}). For example, case A1 refers to 
$\delta m^2=1.75$~eV$^2$, $\sin^22\theta_V=0.10$, and $t_{\rm pb}=1$~s.

\begin{figure}[t]
\centering
    \includegraphics[width=0.99\textwidth]{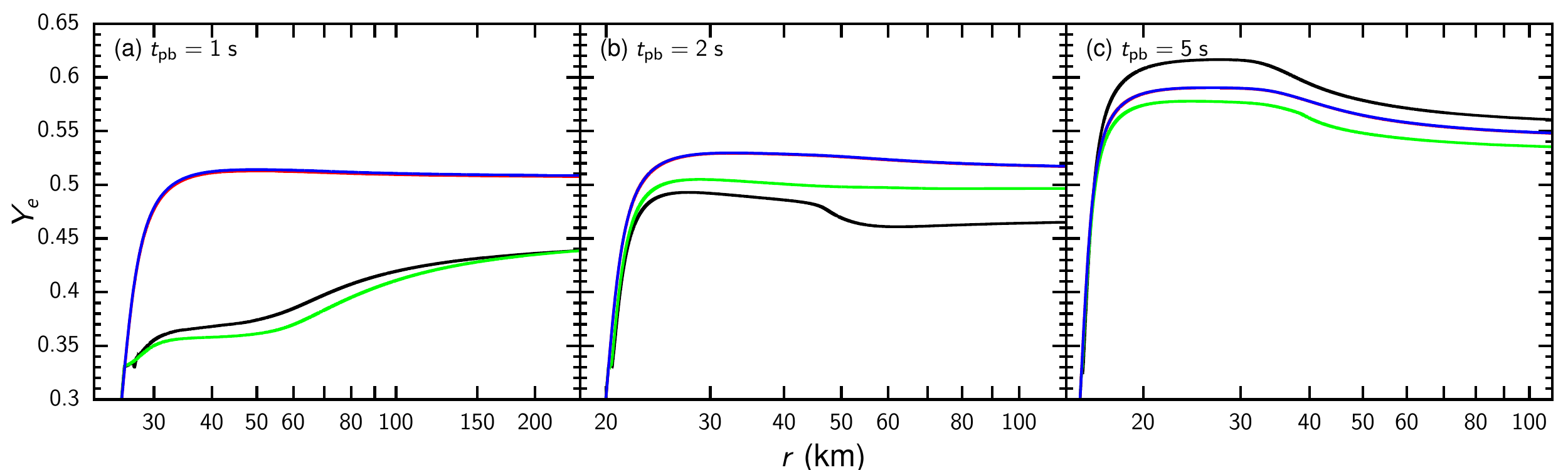}
    \includegraphics[width=0.99\textwidth]{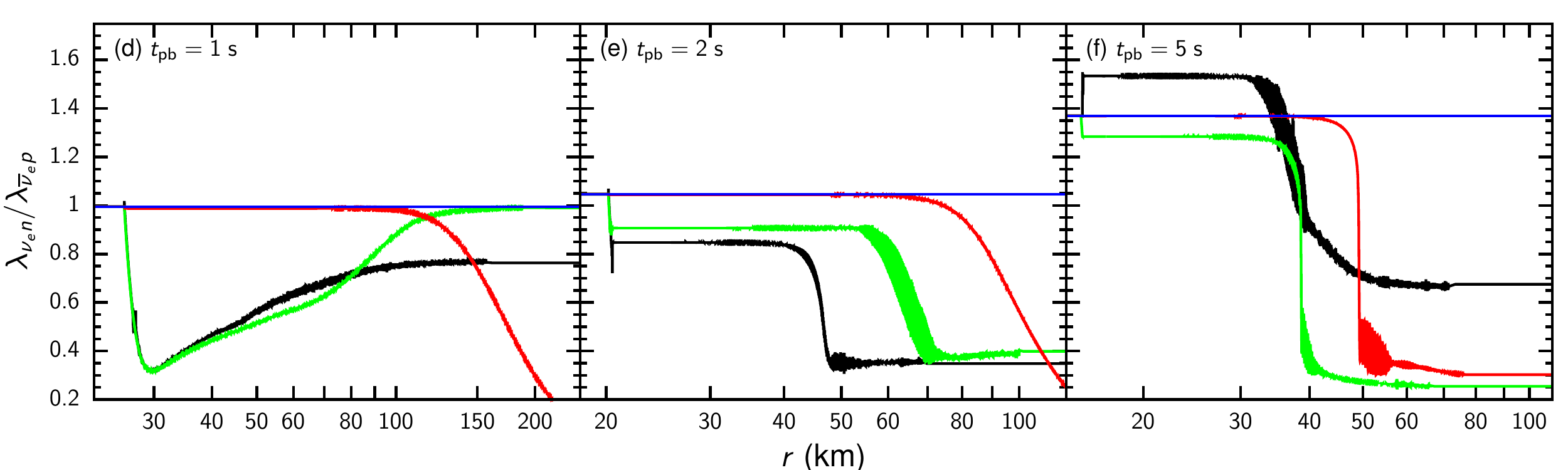}
\caption{\label{fig:YeAll} Electron fraction $Y_e$ (upper row) and ratio of the rates
$\lambda_{\nu_e n}/\lambda_{\bar{\nu}_e p}$ (bottom row) as functions of radius $r$
at $t_{\rm pb}=1$~s (left column), 2~s (central column), and 5~s (right column) 
for cases A (black), B (green), C (red), and D (blue). 
Note that the red and blue curves for $Y_e$ are indistinguishable. 
}
\end{figure}

In Figs.~\ref{fig:YeAll}(a)-(c) we show the $Y_e$ profiles for all the cases.  
Because $\nu_e$ absorption on neutrons increases whereas $\bar\nu_e$ absorption on 
protons decreases $Y_e$, the change of $Y_e$ due to active-sterile neutrino oscillations 
approximately follows the corresponding change of the ratio
$\lambda_{\nu_en}/\lambda_{\bar\nu_ep}$ for the rates of these reactions.
This ratio is shown as a function of $r$ in Figs.~\ref{fig:YeAll}(d)-(f) for all the cases.
Compared with the corresponding cases of no oscillations, cases C1, C2, and C5 show
essentially no change of $\lambda_{\nu_en}/\lambda_{\bar\nu_ep}$, and hence $Y_e$, 
up to $r\sim 90$, 60, and 40~km, respectively. Although 
$\lambda_{\nu_en}/\lambda_{\bar\nu_ep}$ changes greatly at larger radii
for these three cases, there are no corresponding changes of $Y_e$ because $Y_e$ has
essentially frozen out at such large radii due to the small rates of the pertinent
reactions. Therefore, active-sterile neutrino oscillations have negligible effects
on the $Y_e$ profiles in cases C1, C2, and C5. In contrast, there are significant
changes of the $Y_e$ profiles due to oscillations in all the other cases. With the
exception of case A5, active-sterile neutrino oscillations in these cases reduce 
$\lambda_{\nu_en}/\lambda_{\bar\nu_ep}$, and hence $Y_e$, relative to the corresponding 
cases of no oscillations. Case A5 is unique in that oscillations increase
$\lambda_{\nu_en}/\lambda_{\bar\nu_ep}$, and hence $Y_e$. Note again that $Y_e$
eventually freezes out and do not track the effects of oscillations on 
$\lambda_{\nu_en}/\lambda_{\bar\nu_ep}$ at large radii.

\begin{figure}[t]
\centering
    \includegraphics[width=0.99\textwidth]{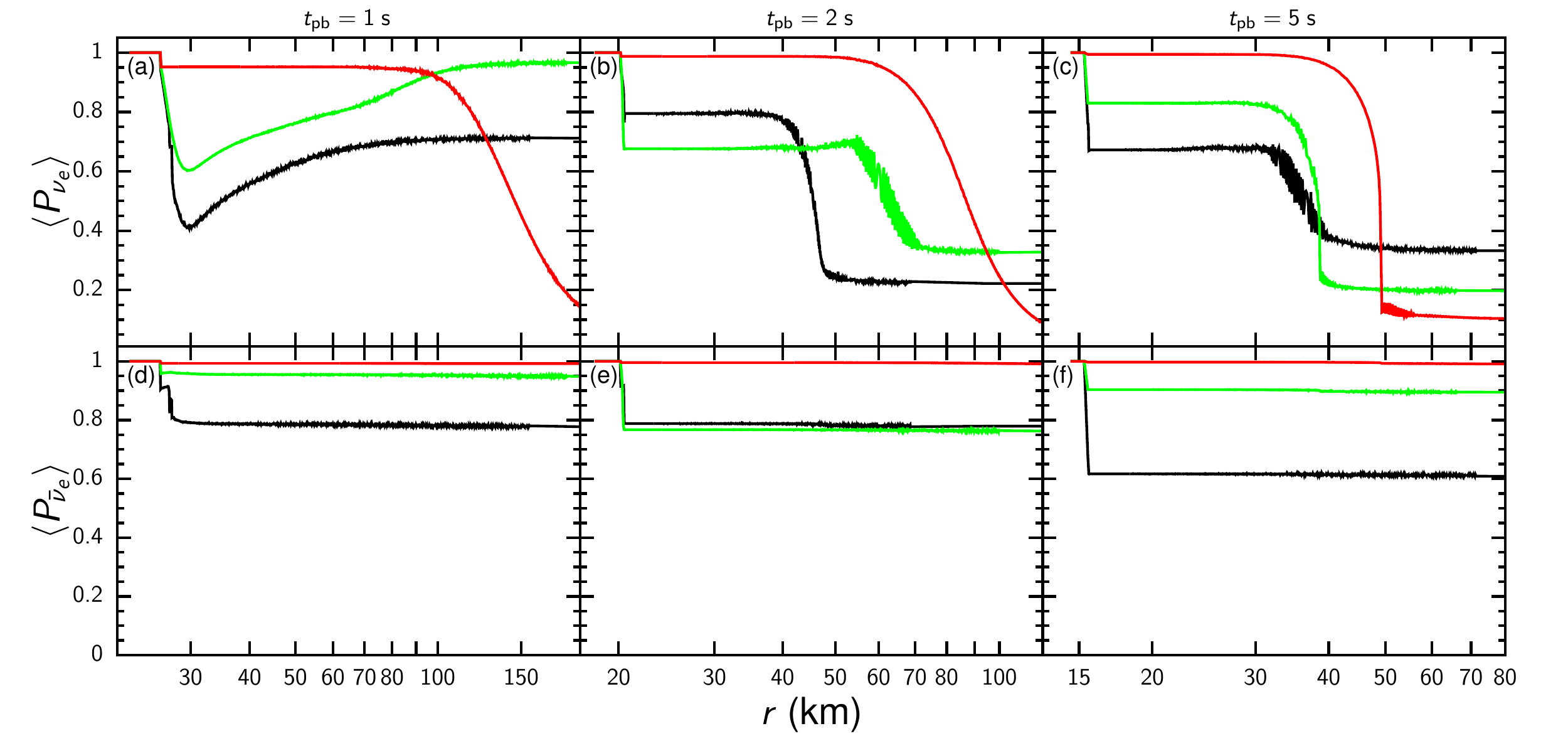}
\caption{\label{fig:probabilityAll} Average survival probabilities for 
$\nu_e$ and $\bar\nu_e$ as functions of radius
at $t_{\rm pb}=1$~s (left column), 2~s (central column), and 5~s (right column) for 
cases A (black), B (green), and C (red). 
}
\end{figure}

To understand the effects of active-sterile neutrino oscillations on $\lambda_{\nu_en}$ 
and $\lambda_{\bar\nu_ep}$, we show in Figure~\ref{fig:probabilityAll} the average survival 
probabilities $\langle P_{\nu_e}(r)\rangle$ and $\langle P_{\bar\nu_e}(r)\rangle$ of $\nu_e$ 
and $\bar\nu_e$, respectively, as functions of $r$ for all the relevant cases. Here for example,
\begin{equation}
    \langle P_{\nu_e}(r)\rangle=\int P_{\nu_e}(E,r)f_{\nu_e}(E)dE,
\end{equation}
where $f_{\nu_e}(E)$ is the normalized $\nu_e$ spectrum at emission
[see Eq.~(\ref{eq:fnu})]. The flavor evolution of $\nu_e$ and $\bar\nu_e$ is sensitive to 
the Mikheyev-Smirnov-Wolfenstein-like (MSW-like) resonances 
\citep{wolfenstein1978neutrino,mikheev1985resonance} corresponding to
\begin{equation}
V_\text{tot}(r)=V_\text{mat}(r)+V_\nu(r)=
\pm \frac{\delta m^2}{ 2 E}\cos 2 \theta_V,
\label{eq:res}
\end{equation}
where the plus and minus signs are for $\nu_e$-$\nu_s$ and $\bar\nu_e$-$\bar\nu_s$
oscillations, respectively. In the above equation, $V_{\rm tot}$ is the total effective 
potential for flavor evolution with contributions $V_{\rm mat}$ [Eq.~(\ref{eq:Vmat})] 
and $V_\nu$ [Eq.~(\ref{eq:Vnu})] from neutrino forward scattering on matter particles 
and other neutrinos, respectively. As a first approximation, the radii for the resonances 
can be estimated from
\begin{equation}
V_\text{mat}(r)=\frac{\sqrt{2}G_F}{2m_u}\rho(r)[3Y_e(r)-1]\sim
\pm\frac{\delta m^2}{2E}\cos2\theta_V.
\label{eq:approxres}
\end{equation}
The term $\rho(3Y_e-1)$ in $V_\text{mat}$ gives rise to two types of resonances. 
The inner resonances occur for both $\nu_e$-$\nu_s$ and $\bar\nu_e$-$\bar\nu_s$ 
oscillations close to the neutrinosphere as $Y_e$ increases from below to slightly 
above 1/3. Here $G_F\rho/m_u$ far exceeds $\delta m^2/E$ and the resonance conditions
are met in the region with $Y_e\sim 1/3$. In contrast, the outer resonances occur only
for $\nu_e$-$\nu_s$ oscillations at much larger radii. Here $Y_e$ is significantly above 
1/3 and the resonance condition is met only for the upper sign in Eq.~(\ref{eq:approxres}) 
due to the much smaller $\rho$. 

\begin{figure}[t]
\centering
    \includegraphics[width=0.95\textwidth]{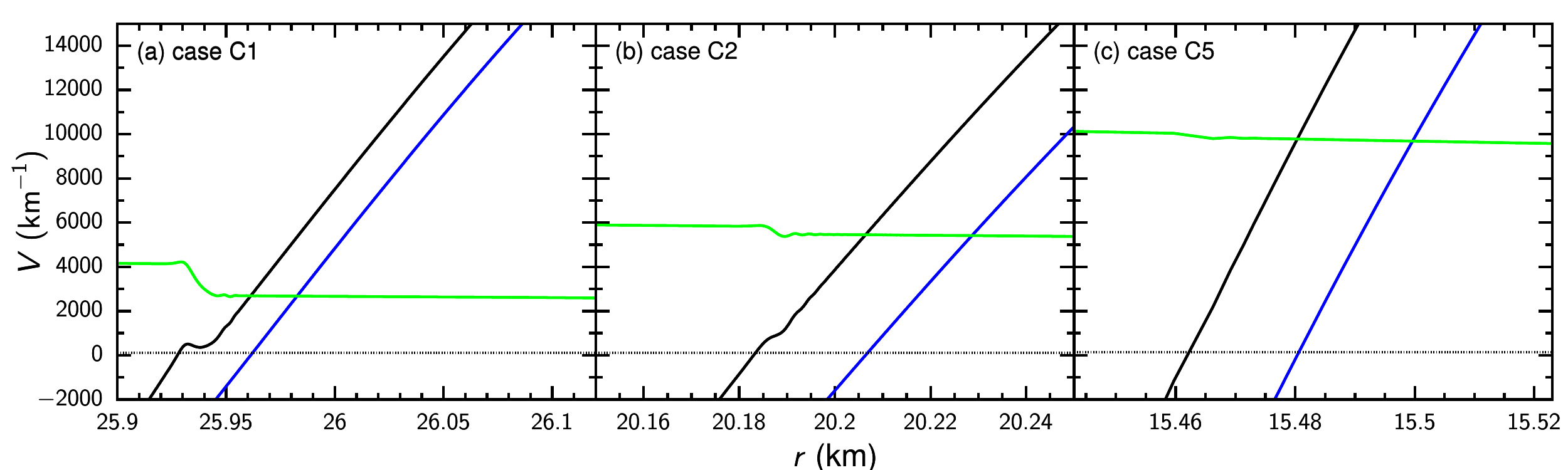}
\caption{\label{fig:VeffAll2} Effective potentials $V_{\rm tot}$ (solid black), $V_{\rm mat}$ 
(solid blue), and $V_\nu$ (solid green) as functions of radius $r$ near the inner resonances
at $t_{\rm pb}=1$~s (a), 2~s (b), and 5~s (c) for case C. For comparison, 
dotted black lines correspond to $\delta m^2/(2\langle E_{\nu_e} \rangle)$.
}
\end{figure}

\begin{figure}[t]
\centering
    \includegraphics[width=0.95\textwidth]{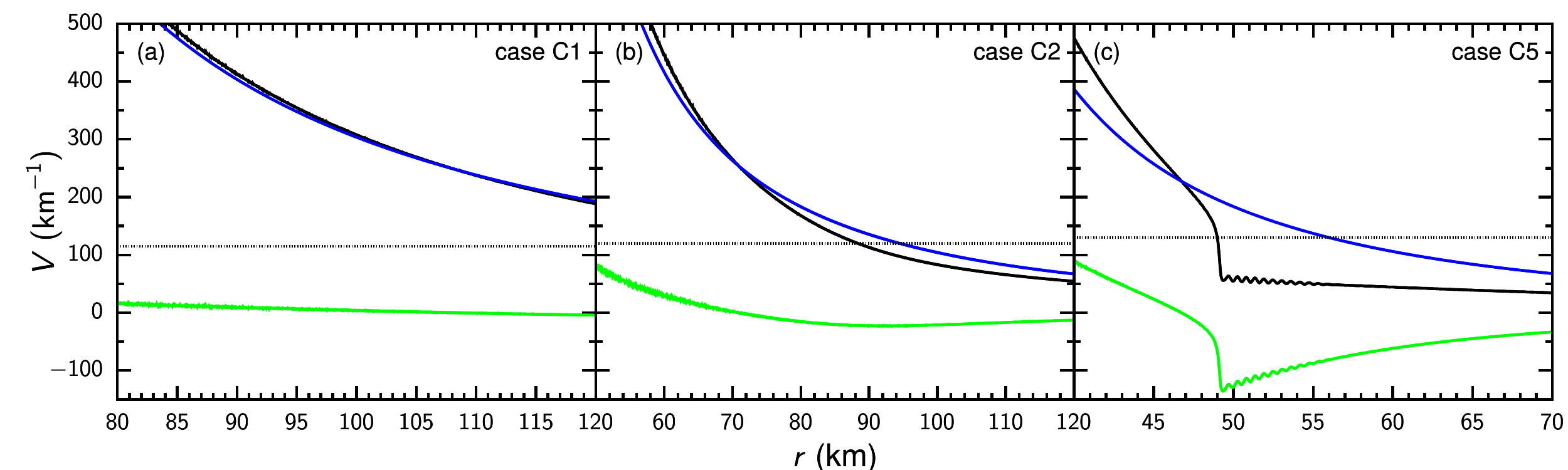}
\caption{\label{fig:VeffAll3} Effective potentials $V_{\rm tot}$ (solid black), $V_{\rm mat}$ 
(solid blue), and $V_\nu$ (solid green) as functions of radius $r$ near the outer resonances
at $t_{\rm pb}=$1~s (a), 2~s (b), and 5~s (c) for case C. For comparison, 
dotted black lines correspond to $\delta m^2/(2\langle E_{\nu_e} \rangle)$.
}
\end{figure}

The main features of oscillations in cases C1, C2, and C5
are the two types of resonances outlined above. Using $V_{\rm tot}$ instead of 
$V_{\rm mat}$ in the resonance conditions does not change the discussion 
qualitatively except for the outer resonance in case C5. 
As shown in Fig.~\ref{fig:VeffAll2}, for a fixed $\nu_e$ or $\bar\nu_e$ energy, 
the inner resonance condition is met at a slightly smaller $r$ when $V_{\rm tot}$ 
is used. With the small values of the corresponding $\delta m^2$ and 
$\sin^22\theta_V$, flavor evolution through the inner resonances is rather 
non-adiabatic for $\nu_e$ and $\bar\nu_e$ of typical energies, 
which leads to large $\langle P_{\nu_e}(r)\rangle$ and $\langle P_{\bar\nu_e}(r)\rangle$ 
immediately following the inner resonances in cases C1, C2, and C5
(see Fig.~\ref{fig:probabilityAll}). Using $V_{\rm tot}$ again changes only slightly
the radius of the outer resonance for a specific $\nu_e$ energy in cases C1 and C2
(see Figs.~\ref{fig:VeffAll3}(a)-(b)). Because $V_{\rm tot}$ changes
much more slowly at large radii, $\nu_e$ flavor evolution through the outer resonances 
is rather adiabatic, which results in large decrease of 
$\langle P_{\nu_e}(r)\rangle$ following the outer resonances in cases C1 and C2
(see Figs.~\ref{fig:probabilityAll}(a)-(b)). The outer resonance in case C5
becomes qualitatively different when $V_{\rm tot}$ is used. Here the magnitude of
$V_\nu$ becomes comparable to that of $V_\text{mat}$ and $V_{\rm tot}$ essentially 
becomes flat (see Fig.~\ref{fig:VeffAll3}(c)). This behavior resembles that of the
matter-neutrino resonances for active-active neutrino oscillations 
(e.g., \citealt{malkus2016symmetric, wu2016physics}), and results in efficient conversion 
of $\nu_e$ with typical energies in case C5 (see Fig.~\ref{fig:probabilityAll}(c)).
However, because the outer resonances occur after the freeze-out of $Y_e$, they have 
little impact on the $Y_e$ profiles. Therefore, active-sterile neutrino oscillations
in cases C1, C2, and C5 only have small effects on all the wind properties, and will
not be discussed any further.

As can be seen clearly in Fig.~\ref{fig:probabilityAll}(b)-(c), the outer resonances 
also occur to further decrease $\langle P_{\nu_e}(r)\rangle$ in cases A2, B2, A5, and B5.
For case B2, the outer resonances 
are similar to those in case C2. For cases A2, A5, and B5, $V_\nu$ becomes significant and 
the behavior of the outer resonances starts to approach that in case C5. However, all outer
resonances occur at large radii corresponding to small rates of the pertinent neutrino
reactions. Consequently, their effects on the $Y_e$ profiles and other wind properties
are very limited. Below we focus on the inner resonances, which dominate the overall 
flavor evolution in cases A1 and B1 as well as produce in general the predominant effects 
of active-sterile neutrino oscillations on all the wind properties including the $Y_e$ 
profiles.

\begin{figure}[t]
\centering
    \includegraphics[width=0.32\textwidth]{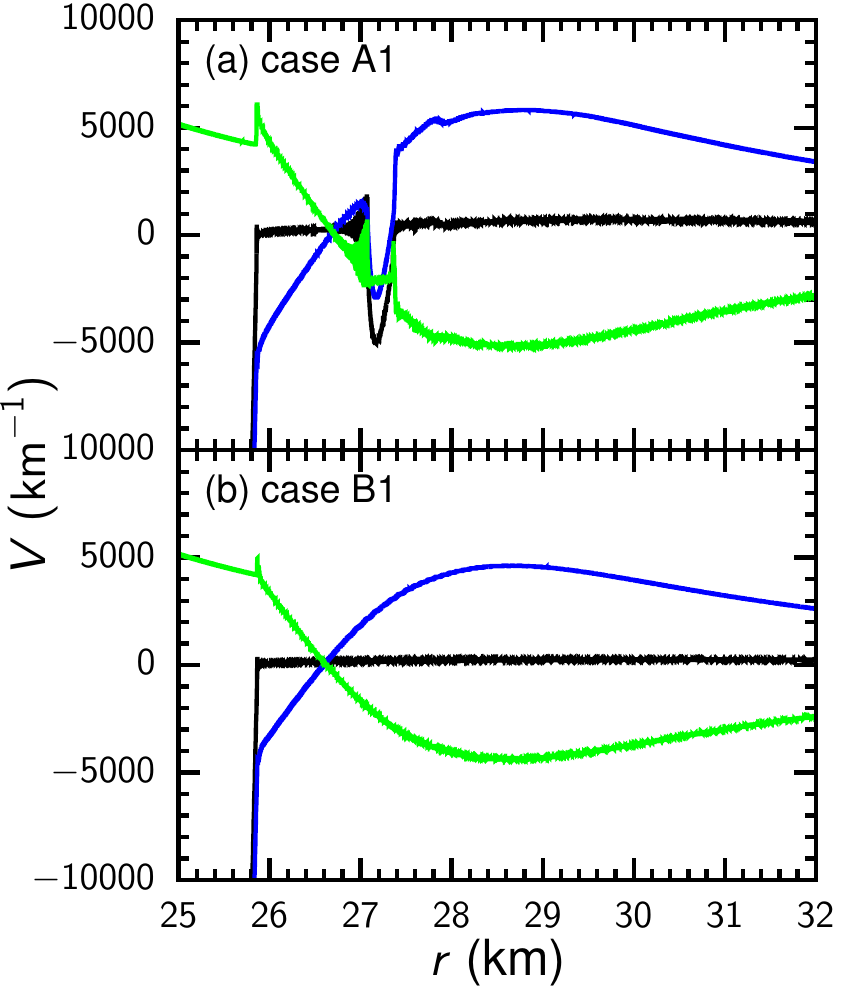}
    \includegraphics[width=0.32\textwidth]{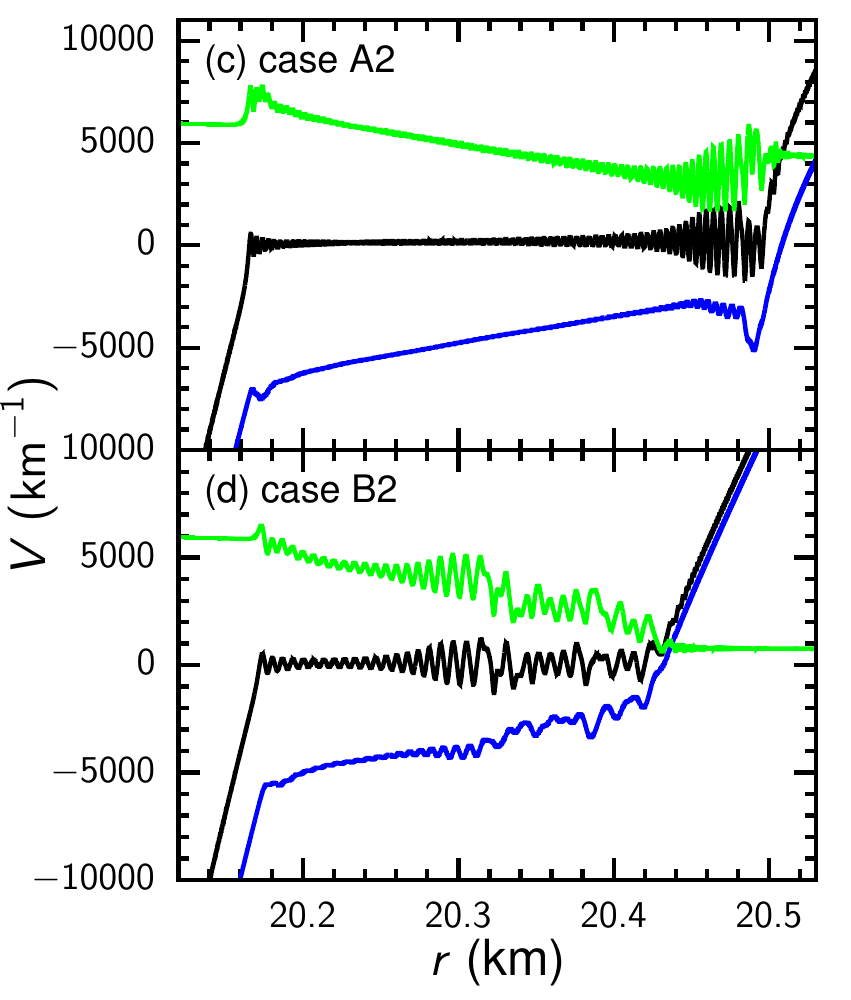}
    \includegraphics[width=0.32\textwidth]{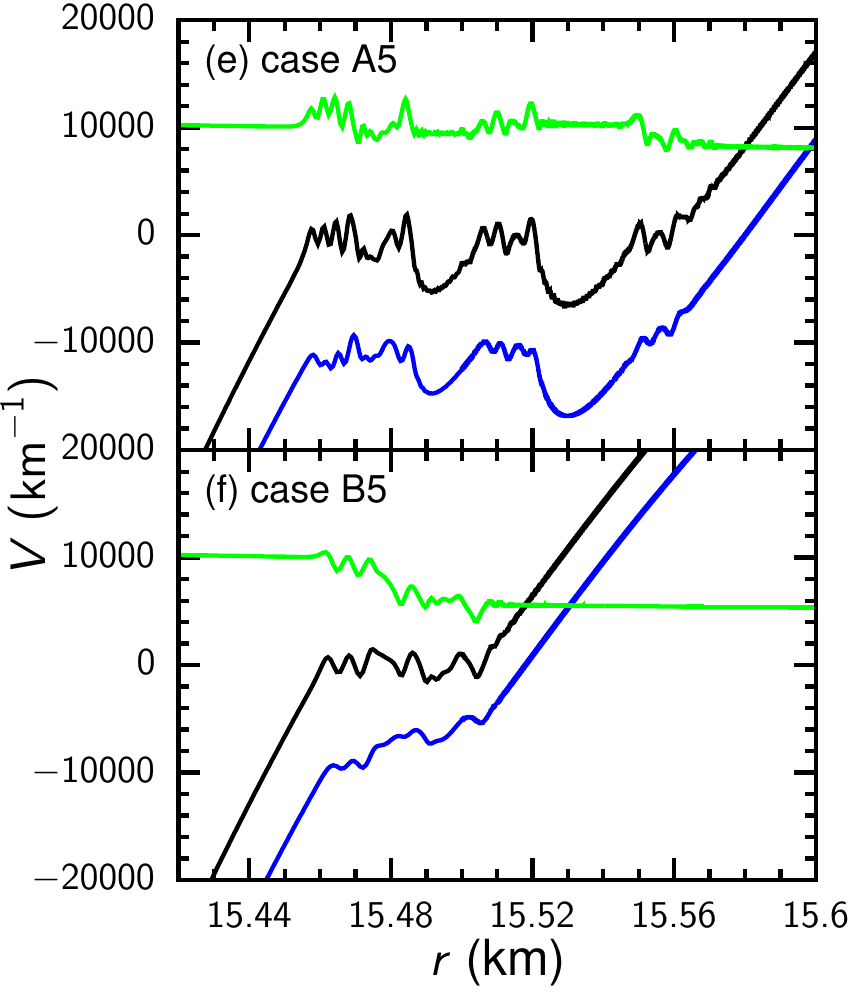}
\caption{\label{fig:VeffAll} Effective potentials $V_{\rm tot}$ (black), $V_{\rm mat}$ 
(blue), and $V_\nu$ (green) as functions of radius $r$ near the inner resonances
at $t_{\rm pb}=1$~s (left column), 2~s (central column), and 5~s (right column) for 
cases A (top row) and B (bottom row). 
}
\end{figure}

\begin{figure}[!hbt]
\centering
    \includegraphics[width=0.49\textwidth]{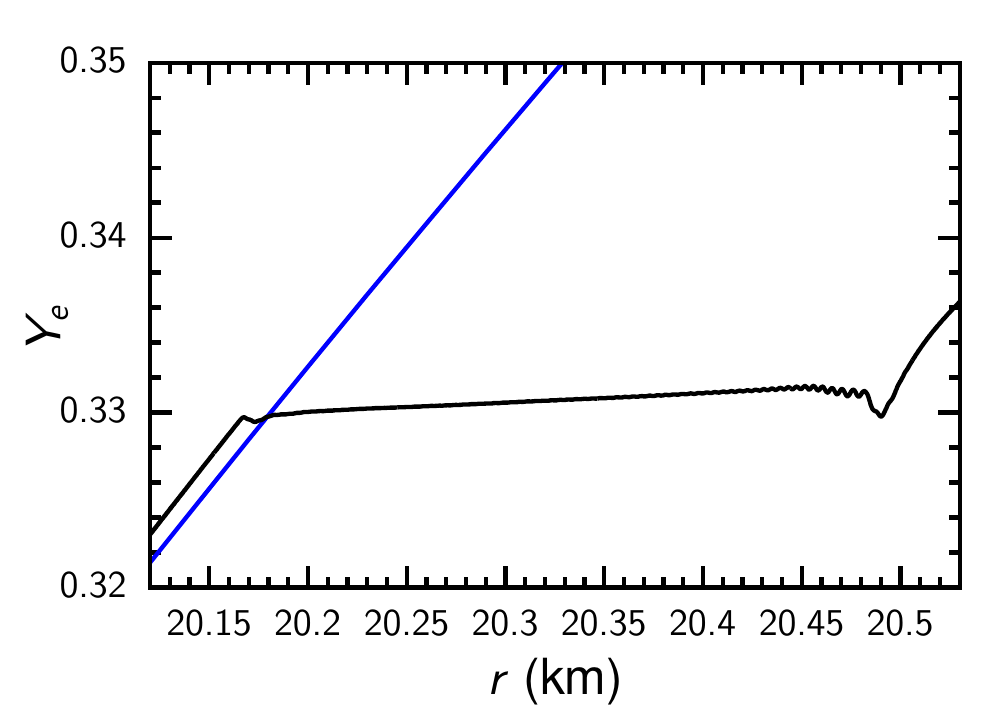}
\caption{\label{fig:Ye2s} Profiles of electron fraction $Y_e$ as functions 
of radius $r$ near the inner 
resonances at $t_{\rm pb}=2$~s for 
cases A (black) and D (blue).
}
\end{figure}

A full understanding of the inner resonances for $\nu_e$-$\nu_s$ and $\bar\nu_e$-$\bar\nu_s$
oscillations requires a careful examination of how the conditions in Eq.~(\ref{eq:res}) are
fulfilled by the contributions $V_{\rm mat}$ and $V_\nu$ to $V_{\rm tot}$. These quantities
are shown as functions of $r$ for cases A1, B1, A2, B2, A5, and B5 in Fig.~\ref{fig:VeffAll}.
All these cases have significant plateaus 
of $V_{\rm tot}\sim\pm\delta m^2\cos2\theta_V/(2E)$ corresponding to inner resonances for
$\nu_e$ and $\bar\nu_e$ of typical energies. These plateaus can be divided into three 
categories: (1) a stable plateau spanning $>1$~km as in case B1, (2) an unstable one spanning
$\sim 0.1$--0.3~km as in cases A2, B2, A5, and B5, and (3) a wide one interrupted
by an instability as in case A1. For all three categories, the plateau results from the near
cancellation of $V_{\rm mat}$ and $V_\nu$, each of which has a magnitude far exceeding that of
$V_{\rm tot}$ for the most part of the corresponding region. The $Y_e$ in this region
is nearly constant and stays close to 1/3. This dramatic flattening of the $Y_e$ profile
corresponding to the inner resonances is shown for case A2 as an example in Fig.~\ref{fig:Ye2s}.

\begin{figure}
\centering
    \includegraphics[width=0.9\textwidth]{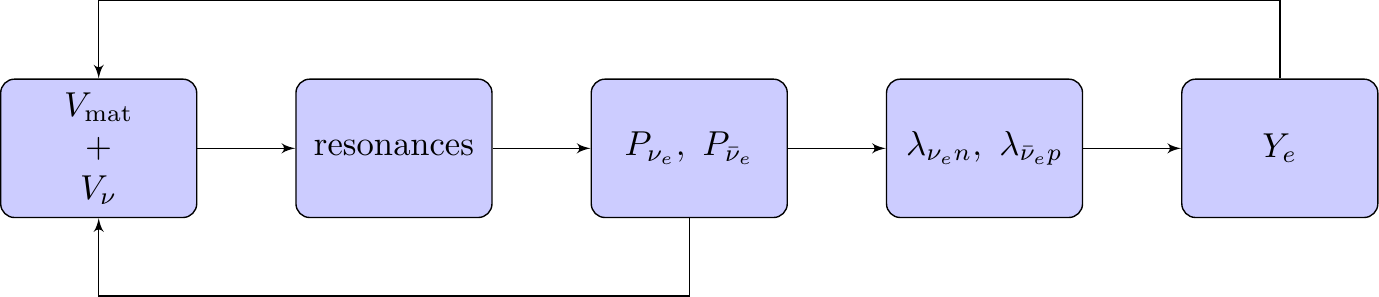}
\caption{\label{fig:flowDiagram} Sketch of the feedback loops for  
active-sterile neutrino oscillations in the wind.
See text for detail.
}
\end{figure}

Formation of the plateau of $V_{\rm tot}$ merits a detailed followup study.
Here we only offer a qualitative sketch of the possible
underlying mechanism as illustrated in Fig.~\ref{fig:flowDiagram}. The total 
potential $V_{\rm tot}$ and the vacuum mixing parameters $\delta m^2$ and
$\sin^22\theta_V$ control the flavor evolution of $\nu_e$ and $\bar\nu_e$, mostly
through the occurrence of resonances. This evolution determines the survival
probabilities $P_{\nu_e}$ and $P_{\bar\nu_e}$, which immediately modify the
contribution $V_\nu$ from neutrino forward scattering on other neutrinos to 
$V_{\rm tot}$. The above factors form the first feedback loop of flavor evolution. 
In addition, $P_{\nu_e}$ and $P_{\bar\nu_e}$ directly modify the rates 
$\lambda_{\nu_en}$ and $\lambda_{\bar\nu_ep}$, which control the evolution of $Y_e$.
In turn, $Y_e$ determines the contribution $V_{\rm mat}$ from neutrino forward 
scattering on matter particles to $V_{\rm tot}$. These factors form the second
feedback loop of flavor evolution. Clearly, the first loop would not operate
were $V_\nu$ far smaller than $V_{\rm mat}$ in magnitude, and the second loop
would not operate if $\lambda_{\nu_en}$ and $\lambda_{\bar\nu_ep}$ were too small
to change $Y_e$ significantly. Consequently, both loops become ineffective
at sufficiently large radii, where $V_\nu$, $\lambda_{\nu_en}$, and 
$\lambda_{\bar\nu_ep}$ are too small due to the geometric dilution of neutrino
fluxes. In contrast, both loops are expected to be efficient close to the
neutrinosphere, thereby forming the plateau of $V_{\rm tot}$ for the inner 
resonances. However, instabilities appear to develop in the feedback loops
under some conditions. Whether and how these instabilities can develop under
different conditions might explain the three categories of plateaus presented
above. We plan to investigate the feedback loops and associated instabilities
in the followup study. We also note that those outer resonances similar to the 
matter-neutrino resonances (e.g., as shown in Fig.~\ref{fig:VeffAll3}(c) for 
case C5) are governed by only the first feedback loop described above.

\section{Results on nucleosynthesis} \label{sec:nucleosynthesis}
In this section, we discuss the impact of active-sterile neutrino oscillations on 
the production of heavy elements in neutrino-driven winds. From our wind models, 
we take a specific set of $v$, $\rho$, and $T$ profiles and use $dt=dr/v$ to define 
the evolution of $\rho(t)$ and $T(t)$ for a mass element as it moves through the wind. 
We use the same established nuclear reaction network as in \citet{Wu:2016pnw} to 
calculate the evolution of the number fraction, or abundance, of each nucleus. 
We start the network calculation at an initial temperature of $T=20$~GK, for which
the nuclear composition is dominated by free nucleons. We take the detailed nuclear 
abundances from the NSE involving all the nuclei in the network for $20\geq T\geq 10$~GK
when NSE holds very well, and evolve the nuclear abundances with the full reaction network 
for $T<10$~GK. Throughout the nucleosynthesis calculations, we follow the evolution of 
$Y_e$ using the rates of $\nu_e$ and $\bar\nu_e$ absorption and $e^\pm$ capture on free 
nucleons as calculated by the wind models. The resulting evolution of $Y_e(t)$ is in
excellent agreement with that calculated directly from the wind models, which always
assume a simple NSE composition with free nucleons and $\alpha$-particles only.
We have also checked that $\dot q_\alpha$, the rate of energy gain per unit mass from 
$\alpha$-particle formation used in the wind models, is a good approximation based on 
the evolution of nuclear abundances in the network calculations. Therefore, the simple 
nuclear composition assumed in the wind models is sufficiently accurate for modeling
the wind dynamics and the evolution of $Y_e$.

For the evolution of $\rho(t)$ and $T(t)$ covered by all our wind models, the $Y_e$ 
values at $T\lesssim 5$~GK are critical to the nucleosynthesis. A more convenient
parameter is the corresponding neutron excess $\delta=1-2Y_e$. For $Y_e\sim 0.5$ with 
$|\delta|\lesssim 0.02$, mainly the Fe group nuclei (mass numbers of $A\sim 56$ and 
atomic numbers of $Z\sim 28$) are produced. For $Y_e>0.5$ with $\delta<-0.02$, the wind
is sufficiently proton-rich and the nuclear flow proceeds beyond the Fe group nuclei 
through the $\nu p$-process \citep{Frohlich:2005ys}. During this process, $\bar\nu_e$ 
absorption on protons can maintain a significant neutron abundance to facilitate 
$(n,p)$ reactions after reactions between charged particles freeze out. In contrast,
for $Y_e<0.5$ with $\delta>0.02$, the wind is sufficiently neutron-rich and the 
nuclear flow proceeds beyond the Fe group nuclei much more efficiently through the 
$\alpha$-process \citep{woosley1992alpha}.

\begin{figure}[t]
\centering
    \includegraphics[width=0.32\textwidth]{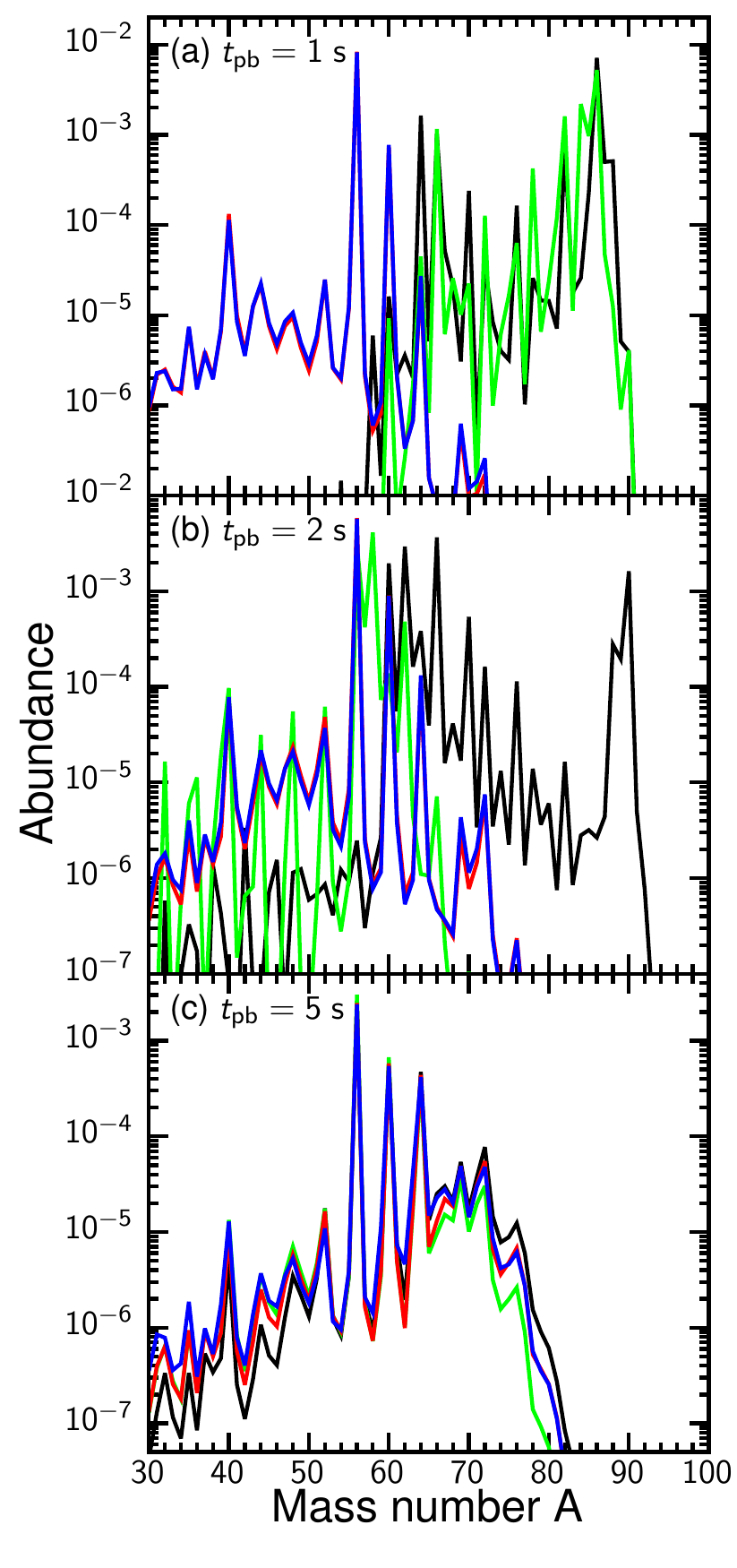}
    \includegraphics[width=0.32\textwidth]{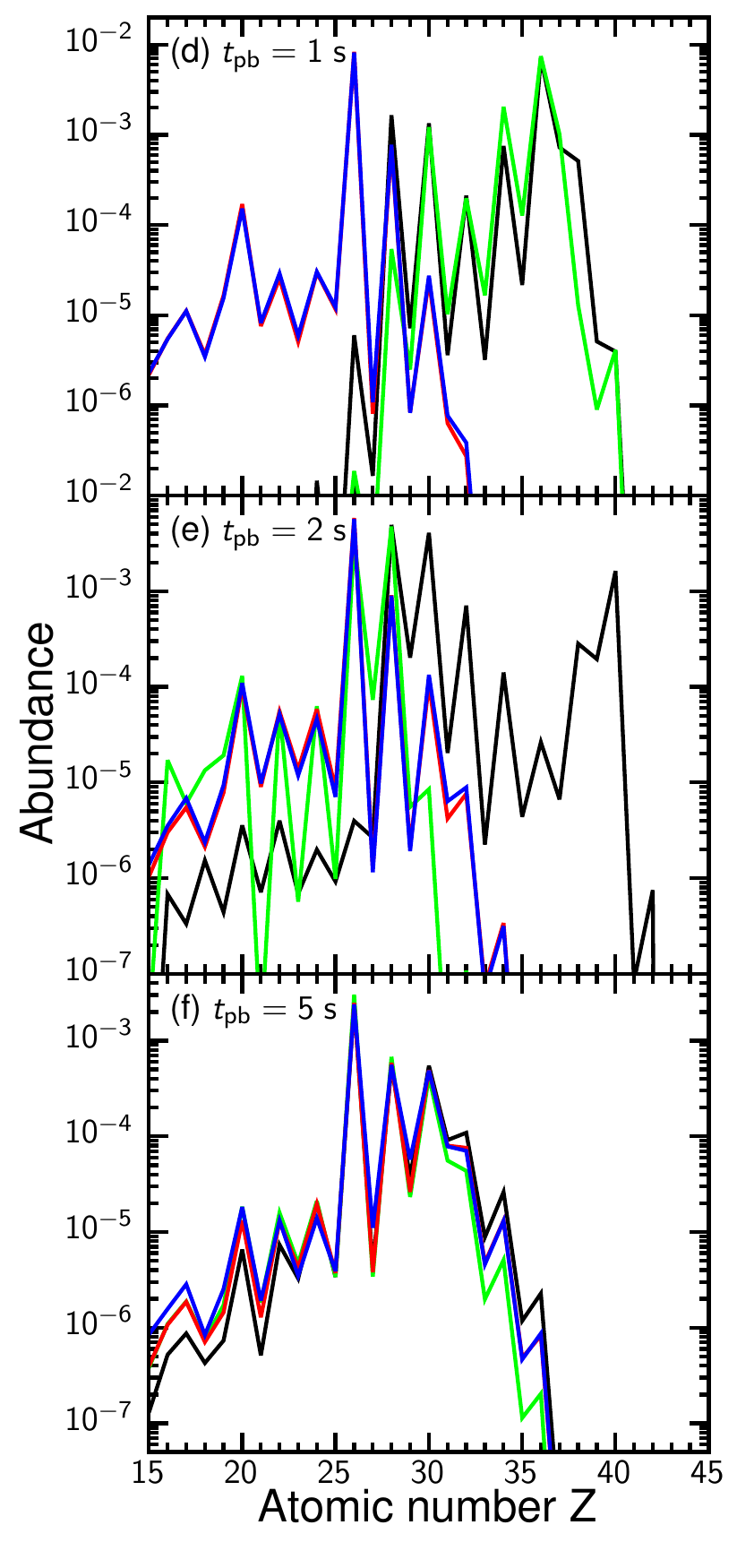}
    \includegraphics[width=0.32\textwidth]{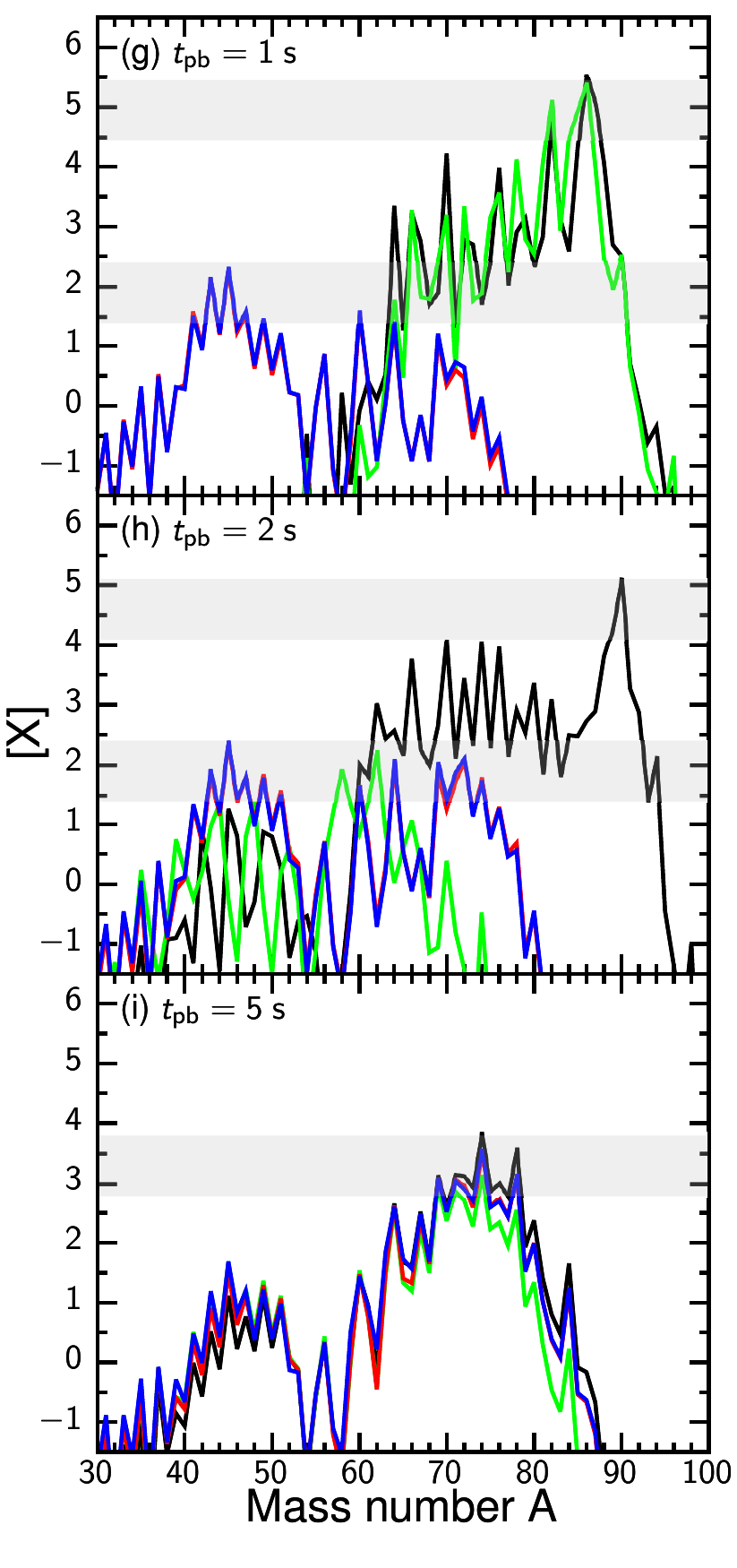}
\caption{\label{fig:abundance1} Abundances as functions of 
mass number (left column) and atomic number (central column), as well as logarithmic  
production factors as functions of mass number (right column)
at $t_{\rm pb}=1$~s (upper row), 2~s (central row), and 5~s (bottom row) 
for cases A (black), B (green), C (red), and D (blue). 
The effective production band is indicated by the grey region in the right column. 
Note that red and blue curves are almost indistinguishable. 
}
\end{figure}

\begin{figure}[t]
\centering
    \includegraphics[width=0.49\textwidth]{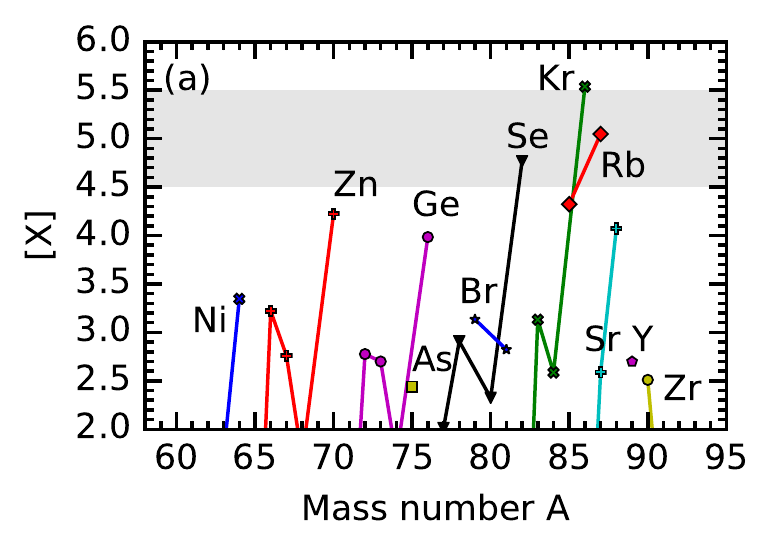}
    \includegraphics[width=0.49\textwidth]{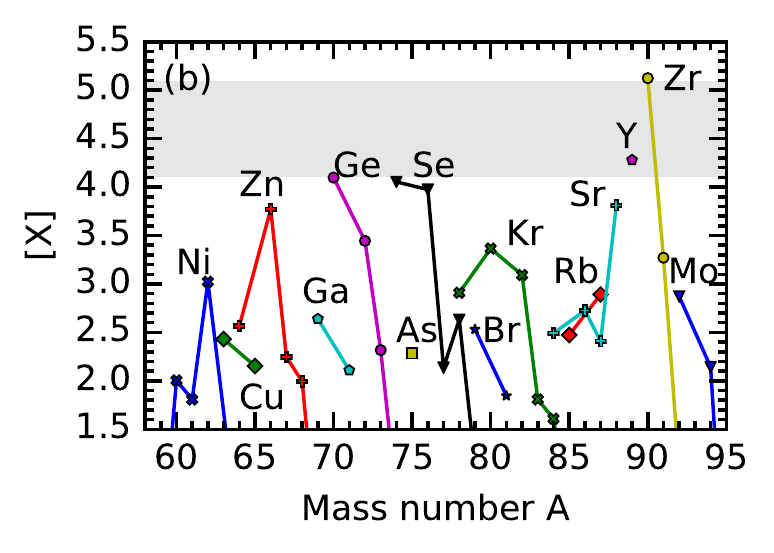}
\caption{\label{fig:abundance2} Production factors as functions of mass number 
for case A at $t_{\rm pb}=1$~s (a) and 2~s (b).
Isotopes of the same element are connected by line segments.
The effective production band is indicated by the grey region.
}
\end{figure}

Figure~\ref{fig:abundance1} shows the final abundances produced in all our wind
models as functions of mass number $A$ (left panels) and atomic number $Z$
(middle panels), respectively. 
In the absence of neutrino oscillations, the wind is generally proton-rich
and becomes more so for the later epochs of the CCSN evolution, with typical values of
$Y_e\sim 0.51$, 0.52, and 0.55 for cases D1, D2, and D5, respectively 
(see Fig.~\ref{fig:YeAll}). Consequently, nucleosynthesis shifts from
dominant production of Fe group nuclei for case D1 to increasing production of heavier 
nuclei through the $\nu p$-process for cases D2 and D5. This trend is more clearly
shown in the right panels of Fig.~\ref{fig:abundance1} in terms of the logarithmic 
production factor $[X]=\log(X/X_\odot)$, where $X$ is the mass fraction of a nucleus 
produced in the wind and $X_\odot$ is the corresponding value in the solar system.
For understanding how the solar composition is obtained by mixing the nucleosynthesis 
contributions from various astrophysical sources, values of $[X]$ are more important than 
the abundances produced by a source. For example, $^{45}$Sc has the highest value of 
$[X]\approx 2.3$ among all the nuclei produced by the wind in case D1, while
$^{56}$Fe has the highest abundance but a much smaller value of $[X]\approx 0.9$.
This result means that even if all the $^{45}$Sc in the solar system came from sources 
like this wind, these sources could have contributed only $\approx 4$\% of the solar
$^{56}$Fe. The horizontal bands in Figs.~\ref{fig:abundance1}-\ref{fig:abundance2}
indicate $[X]$ within 1~dex of the highest value. Nuclei within these bands
would be the most relevant for contributions to the solar composition.

Because active-sterile neutrino oscillations have little impact on the $Y_e$ profiles in 
cases C1, C2, and C5, nucleosynthesis in these cases is essentially identical to that in 
the corresponding cases of no oscillations. In addition, all cases have similar 
nucleosynthesis for $t_{\rm pb}=5$~s because the corresponding winds are similarly 
proton-rich (see Fig.~\ref{fig:abundance1}). In contrast, 
oscillations greatly reduce the $Y_e$ values in cases A1, B1, and A2 
(see Fig.~\ref{fig:YeAll}), and dramatically alter the nucleosynthesis to produce
nuclei far beyond the Fe group (see Fig.~\ref{fig:abundance1}). 
The largest $[X]$ is $\approx 5.5$ and 5.4 for $^{86}$Kr in cases A1 and B1, respectively,
and is $\approx 5.1$ for $^{90}$Zr in case A2 (see Fig.~\ref{fig:abundance2}). Both nuclei
have a magic number (50) of neutrons and are produced by the $\alpha$-process.
Finally, oscillations also reduce $Y_e$ significantly in case B2, but the resulting
values of $Y_e\sim 0.5$ correspond to a very small $|\delta|$. Consequently,
case B2 has the least extensive production of nuclei beyond the Fe group among all the 
cases for $t_{\rm pb}=2$~s. Specifically, $^{62}$Ni and $^{58}$Ni have
the largest two $[X]$ values of $\approx 2.2$ and 1.9, respectively, in this case, 
while $^{45}$Sc and $^{72}$Ge have the largest two $[X]$ values of $\approx 2.4$ and
2.1, respectively, in case D2.

\section{Conclusions} \label{sec:conclusions}
We have included active-sterile neutrino oscillations in a steady-state model of
the neutrino-driven wind and self-consistently treated the effects of oscillations
on the $v$, $\rho$, $T$, and $Y_e$ profiles of the wind. Compared to previous studies
(e.g., \citealt{nunokawa1997supernova,mclaughlin1999active,tamborra2012impact,%
wu2014impact,pllumbi2015impact})
that addressed only the effects on the $Y_e$ profile, our study represents
a significant step forward. Among the three sets of vacuum mixing parameters adopted
here (see Table~\ref{tab:NDWnumericalOscillationParameters}), 
we find that only those of case C
with $\delta m^2=0.4$~eV$^2$ produce negligible effects on the wind. In contrast,
those of cases A and B with $\delta m^2=1.75$ and 1.0~eV$^2$, respectively, 
significantly change the wind dynamics in addition to the $Y_e$ profile. Specifically, 
the mass loss rate is reduced by a factor of $\approx 2.7$, 1.6, and 1.6 for 
$t_{\rm pb}=1$, 2, and 5~s in cases A1, A2, and A5, respectively 
(see Table~\ref{tab:NDWnumericalMdot}). This reduction translates into
a similar reduction of the wind velocity and is caused by the reduced heating 
as $\nu_e$ and $\bar\nu_e$ are converted into their sterile counterparts.
While the effects of oscillations on the $\rho$ and $T$ profiles are much smaller,
the resulting increases in $S_{\rm tot}$ by $\sim 15\%$, 5\%, and 5\% in cases A1, A2, 
and A5, respectively (see Fig.~\ref{fig:EntropyAll}), are still noticeable.

With respect to nucleosynthesis, the most important effects of active-sterile neutrino 
oscillations are the resulting changes of $Y_e$, which range from significant
in cases B2, A5, and B5 to large in cases A1, B1, and A2 (see Fig.~\ref{fig:YeAll}).
With the large reduction of $Y_e$ in the latter three cases, nucleosynthesis is
dramatically altered (see Figs.~\ref{fig:abundance1}-\ref{fig:abundance2}). 
For cases D1 and D2 with no oscillations, the final abundance pattern is characterized 
by the largest $[X]\approx 2.3$ and 2.4, respectively, for $^{45}$Sc. In contrast,
$^{86}$Kr is produced with the largest $[X]\approx 5.5$ and 5.4 in cases A1 and B1,
respectively, and $^{90}$Zr is produced with the largest $[X]\approx 5.1$ in case A2.
Not only do oscillations change the dominant nuclei produced in these cases,
but they also greatly amplify the potential contributions from the corresponding 
winds to the solar composition. The effects of oscillations are also large but less 
dramatic in case B2, where $^{62}$Ni is produced with the largest $[X]\approx 2.2$
(see Fig.~\ref{fig:abundance1}). On the other hand, oscillations do not change the
nucleosynthesis very much in cases A5 and B5 despite the significant changes of
$Y_e$ (see Fig.~\ref{fig:abundance1}). This result comes about because the $Y_e$
in these cases corresponds to a proton-rich wind that undergoes the $\nu p$-process
similar to the case of no oscillations. We conclude that active-sterile neutrino
oscillations with vacuum mixing parameters similar to those in cases A and B can 
greatly affect nucleosynthesis in the winds at the early epochs of 
$t_{\rm pb}\sim 1$--2~s during the CCSN evolution.

In agreement with previous studies (e.g., \citealt{wu2014impact}),
we find that active-sterile neutrino oscillations
near the PNS exhibit features beyond the usual MSW effect that is dominated by the 
potential $V_{\rm mat}$ from neutrino forward scattering on matter particles. These
interesting features are caused by the potential $V_\nu$ from neutrino forward 
scattering on other neutrinos, and more importantly, by the feedback of oscillations 
on both $V_{\rm mat}$ and $V_\nu$ (see Fig.~\ref{fig:flowDiagram}). In particular,
we have observed plateau-like behaviors of $V_{\rm tot}=V_{\rm mat}+V_\nu$ for the 
outer MSW-like resonances in case C5 (see Fig.~\ref{fig:VeffAll3}(c)) and for the 
inner resonances in all the wind models with vacuum mixing parameters of cases A and B 
(see Fig.~\ref{fig:VeffAll}). We have given only a qualitative explanation 
of such behaviors here, and plan to investigate them in detail in a followup study.
In addition, we have made some simplifying approximations in our present study. 
For example, we have assumed that all active neutrinos are emitted from the same sharp 
neutrinosphere. In reality, where neutrinos decouple from matter depends on the neutrino 
species and energy. Although using slightly different neutrinospheres for $\nu_e$, 
$\bar\nu_e$, and $\nu_x$ ($\bar\nu_x$) is more realistic, we do not expect that this 
modification would affect our results qualitatively. Nevertheless, we also plan to check 
the effects of this and other possible improvements of our model in the followup study.

\acknowledgments
We thank Gabriel Mart\'inez-Pinedo for discussion during the initial
phase of our study. We also acknowledge the Minnesota Supercomputing 
Institute at the University of Minnesota for providing resources 
that contributed to the research results reported here. 
This work was supported in part by the US Department of Energy 
[DE-FG02-87ER40328 (UM)], the Ministry of Science and Technology 
of Taiwan [107-2119-M-001-038 (IOP)], the National Natural Science 
Foundation of China [11533006, 11655002 (TDLI)], and the Science and 
Technology Commission of Shanghai Municipality [16DZ2260200 (TDLI)].

\appendix

\section{Rates of neutron-proton interconversion and of energy gains and losses}
\label{sec:Allrates}

The rate $\lambda_{\nu_en}$ is given by Eq.~(\ref{eq:lambdanuen}).
Using the same notation, we can write $\lambda_{\bar\nu_ep}$ as
\begin{equation}
    \lambda_{\bar{\nu}_e p} = \frac{G_F^2 |V_{ud}|^2 (1+3 g_A^2)}{2\pi^2} 
    \frac{L_{\bar\nu_e} D(r)}{R^2_\nu \langle E_{\bar\nu_e} \rangle} 
    \int_{E_{\text{th}}}^\infty (E-\Delta)^2\left(1-\frac{W_{\bar{\nu}_e}E}{m_N}\right)
    f_{\bar\nu_e}(E)P_{\bar\nu_e}(E,r)dE,
\end{equation}
where $E_{\rm th}=\Delta+m_e$ is the threshold $\bar\nu_e$ energy for absorption on
protons, $m_e$ is the electron rest mass, and
\begin{equation}
	W_{\bar{\nu}_e} = \frac{2[1 + 5 g_A^2 + 2 g_A (1 + f_2)]}{1 + 3 g_A^2}.
\end{equation}
In the above equations, $V_{ud}=0.974$, $g_A=1.27$, and $f_2=3.71$.
The rates $\lambda_{e^+ n}$ and $\lambda_{e^-p}$ are
\begin{alignat}{2}
    	\lambda_{e^+ n} &= \frac{G_F^2 |V_{ud}|^2 (1+3 g_A^2)}{2\pi^3} T^5 
    	\left[ g_{0,2,0}\left(\frac{m_e}{T},\eta,\frac{\Delta}{T},0\right) 
    	- \frac{W_{\bar{\nu}_e} T}{m_N} 
    	g_{1,2,0}\left(\frac{m_e}{T},\eta, \frac{\Delta}{T}, 0\right) \right],\\
    	\lambda_{e^-p} &= \frac{G_F^2 |V_{ud}|^2 (1+3 g_A^2)}{2\pi^3} T^5 
    	\left[ g_{0,2,0}\left(\frac{\Delta}{T},-\eta,-\frac{\Delta}{T},0\right) 
    	- \frac{W_{\nu_e} T}{m_N} 
    	g_{1,2,0}\left(\frac{\Delta}{T},-\eta, -\frac{\Delta}{T}, 0\right) \right],
\end{alignat}
respectively, where $\eta=\mu/T$, $\mu$ is the electron chemical potential, and
\begin{equation}
    g_{i,j,k}(x_{\rm th},y,z,w)=\int_{x_\text{th}}^\infty 
    \frac{x^{i+1}(x+z)^j(x+w)^k\sqrt{x^2-(m_e/T)^2}}{\exp(x+y)+1}dx.
\end{equation}

The rates $\dot{q}_{\nu_e n}$ and $\dot{q}_{\bar{\nu}_e p}$ are
\begin{alignat}{2}
    	\dot{q}_{\nu_e n} &= \frac{Y_n}{m_u}\frac{G_F^2 |V_{ud}|^2 (1+3 g_A^2)}{2\pi^2} 
    	\frac{L_{\nu_e} D(r)}{R^2_\nu \langle E_{\nu_e} \rangle}
        \int_0^\infty[E+(\Delta - m_e)](E+\Delta)^2 
        \left(1- \frac{W_\nu E}{m_N}\right)f_{\nu_e}(E)P_{\nu_e}(E,r)dE,\\
        \dot{q}_{\bar\nu_e p} &= \frac{Y_p}{m_u}\frac{G_F^2 |V_{ud}|^2 (1+3 g_A^2)}{2\pi^2} 
    	\frac{L_{\bar\nu_e} D(r)}{R^2_\nu \langle E_{\bar\nu_e} \rangle}
        \int_{E_{\rm th}}^\infty[E-(\Delta - m_e)](E-\Delta)^2 
        \left(1- \frac{W_{\bar\nu} E}{m_N}\right)f_{\bar\nu_e}(E)P_{\bar\nu_e}(E,r)dE,
\end{alignat}
respectively. The rates $\dot{q}_{e^+ n}$ and $\dot{q}_{e^- p}$ are
\begin{alignat}{2}
    	\dot{q}_{e^+ n} &=\frac{Y_n}{m_u}\frac{G_F^2 |V_{ud}|^2 (1+3 g_A^2)}{2\pi^3} T^6
        \left[ g_{0,2,1}\left(\frac{m_e}{T},\eta,\frac{\Delta}{T},\frac{m_e}{T}\right) - 
        \frac{W_{\bar{\nu}_e} T}{m_N} 
        g_{1,2,1}\left(\frac{m_e}{T},\eta,\frac{\Delta}{T},\frac{m_e}{T}\right) \right],\\
        \dot{q}_{e^-p} &=\frac{Y_p}{m_u}\frac{G_F^2 |V_{ud}|^2 (1+3 g_A^2)}{2\pi^3} T^6
        \left[ g_{0,2,1}\left(\frac{\Delta}{T},-\eta,-\frac{\Delta}{T},-\frac{m_e}{T}\right) - 
        \frac{W_{\nu_e} T}{m_N} 
        g_{1,2,1}\left(\frac{\Delta}{T},-\eta,-\frac{\Delta}{T},-\frac{m_e}{T}\right) \right],
\end{alignat}
respectively.

The rates given above are the most important for modeling the neutrino-driven wind, 
and therefore, are calculated with a higher accuracy. The other rates for energy gains
and losses are less important. We adopt similar estimates of these rates to those in
\cite{thompson2001physics} but ignore the effects of general relativity. These
approximate rates are given below.

The rate $\dot{q}_{\nu_e N}$ is
\begin{equation}
    \dot{q}_{\nu_e N} = \frac{\kappa_nY_n+\kappa_pY_p}{m_u^2} \frac{G_F^2}{2\pi^2} 
    \frac{L_{\nu_e} D(r)}{R_\nu^2 \langle E_{\nu_e} \rangle}
    \int_0^\infty E^3(E-6T)f_{\nu_e}(E)P_{\nu_e}(E,r)dE, 
\end{equation} 
where $\kappa_n=\frac{1}{4}(1+3g_A^2)$, 
$\kappa_p=(\frac{1}{2}-2\sin^2\theta_W)^2+\frac{3}{4}g_A^2$,
and $\sin^2\theta_W=0.231$. The above rate can be adapted in a straightforward manner
to give $\dot{q}_{\bar\nu_e N}$, $\dot{q}_{\nu_x N}$, and $\dot{q}_{\bar\nu_x N}$.
Note that all active neutrinos and antineutrinos of the same energy have the same cross 
sections for neutral-current scattering on nucleons.

The rate $\dot{q}_{\nu_e e^\pm}$ is
\begin{equation}
    \dot{q}_{\nu_e e^\pm} = 
    [\Lambda_{\nu_e e^-}n_{e^-}+\Lambda_{\nu_e e^+}n_{e^+}]
    \frac{G_F^2}{2\pi^2}\frac{T}{\rho}
    \frac{L_{\nu_e} D(r)}{R_\nu^2 \langle E_{\nu_e} \rangle} 
    \int_0^\infty E(E-4T) f_{\nu_e}(E) P_{\nu_e}(E,r) dE, 
\end{equation}
where $n_{e^\pm}=(T^3/\pi^2)g_{0,0,0}(m_e/T,\pm\eta,0,0)$ are 
the number densities of positrons and electrons, respectively,
$\Lambda_{\nu_e e^-}=(C_V+C_A)^2+\frac{1}{3}(C_V-C_A)^2$,
$\Lambda_{\nu_e e^+}=(C_V-C_A)^2+\frac{1}{3}(C_V+C_A)^2$, 
$C_V=\frac{1}{2}+2\sin^2\theta_W$, and $C_A=\frac{1}{2}$.
In adapting the above rate to give 
$\dot{q}_{\bar\nu_e e^\pm}$, $\dot{q}_{\nu_x e^\pm}$, and 
$\dot{q}_{\bar\nu_x e^\pm}$, note that
$\Lambda_{\bar\nu_e e^-}=\Lambda_{\nu_e e^+}$,
$\Lambda_{\bar\nu_e e^+}=\Lambda_{\nu_e e^-}$,
$\Lambda_{\nu_x e^-}=(C_V+C_A-2)^2+\frac{1}{3}(C_V-C_A)^2$,
$\Lambda_{\nu_x e^+}=(C_V-C_A)^2+\frac{1}{3}(C_V+C_A-2)^2$,
$\Lambda_{\bar\nu_x e^-}=\Lambda_{\nu_x e^+}$, and
$\Lambda_{\bar\nu_x e^+}=\Lambda_{\nu_x e^-}$.

The rate $\dot{q}_{\nu_e \bar\nu_e}$ is
\begin{equation}
    \dot{q}_{\nu_e \bar\nu_e} = 
    \frac{\Lambda_{\nu_e \bar\nu_e}}{\rho} \frac{G_F^2}{36\pi^3} 
    \frac{L_{\nu_e} L_{\bar\nu_e} \Psi(r)}{R_\nu^4 \langle E_{\nu_e} 
    \rangle \langle E_{\bar\nu_e} \rangle} 
    \int_0^\infty \int_0^\infty E E' (E+E') 
    f_{\nu_e}(E) P_{\nu_e}(E,r)
    f_{\bar\nu_e}(E') P_{\bar\nu_e}(E',r)dEdE', 
\end{equation}
where $\Psi(r)=[D(r)]^4\{[D(r)]^2-6D(r)+10\}$, 
$\Lambda_{\nu_e \bar\nu_e}=C_V^2+C_A^2$.
In adapting the above rate to give $\dot{q}_{\nu_x \bar\nu_x}$,
note that $\Lambda_{\nu_x \bar\nu_x}=(C_V-1)^2+(C_A-1)^2$.
The rate $\dot{q}_{e^- e^+}$ is
\begin{equation}
    \dot{q}_{e^- e^+} = 
    (\Lambda_{\nu_e \bar\nu_e} + 2\Lambda_{\nu_x \bar\nu_x})
    \frac{2G_F^2}{9\pi^5} \frac{T^9}{\rho} 
    \left[g_{1,0,0}\left(\frac{m_e}{T},\eta,0,0\right)
    g_{2,0,0}\left(\frac{m_e}{T},-\eta,0,0\right)+
    g_{1,0,0}\left(\frac{m_e}{T},-\eta,0,0\right)
    g_{2,0,0}\left(\frac{m_e}{T},\eta,0,0\right)\right].
\end{equation}

\bibliographystyle{apj}
\bibliography{references}

\end{document}